\documentclass{aa}  

\usepackage{graphicx}
\usepackage{txfonts}
\usepackage{tabularray}
\usepackage{afterpage}
\usepackage{float}
\usepackage{txfonts}
\usepackage[version=4]{mhchem}
\usepackage{xcolor}

\newcommand{\kjmol}{kJ mol$^{-1}$\,}

\usepackage{longtable}
\usepackage{booktabs}
\usepackage{array}
\usepackage{rotating}

\newif\ifrevision
\revisionfalse 

\makeatletter
\newcounter{rxn}
\newcommand{\rxnlabel}[2]{%
  \refstepcounter{rxn}%
  \def\@currentlabel{#2}%
  \textbf{#2}\label{#1}%
}
\makeatother

\begin{document}

\title{Methanimine as a sink in the HCN and HNC solid state hydrogenation network}
\titlerunning{Methanimine as a sink in the HCN and HNC solid state hydrogenation network}

\author{J. Enrique-Romero*
          \inst{1}
          \, T. Lamberts\inst{1,2}
          }

\institute{%
Leiden Institute of Chemistry, Gorlaeus Laboratories, Leiden University, PO Box 9502, 2300 RA Leiden, The Netherlands\\
\email{j.enrique.romero@lic.leidenuniv.nl}
\and
Leiden Observatory, Leiden University, P.O. Box 9513, 2300 RA Leiden, The Netherlands
}

   \date{}
 
  \abstract
    {}
   {We aim to provide a systematic and quantitative description of the hydrogenation network connecting HCN and HNC to methylamine on interstellar water ices, and to identify the dominant reaction pathways and potential bottlenecks.}
   {We performed a comprehensive quantum-chemical investigation of H-addition, H-abstraction, reactions with \ce{H2}, and water-assisted H-transfer isomerization involving all intermediates connecting \ce{HCN} and \ce{HNC} up to \ce{CH3NH2}, considering amorphous solid water molecular clusters composed of 14 water molecules. We employed benchmarked density functional theory to derive activation barriers, elucidate the reaction mechanisms, and determine the binding energy distribution of \ce{H2CN} and \ce{CNH2}. We also discuss the effect of deuterium substitutions on selected points of the network.}
    {H-addition reactions are generally mediated by activation energy barriers, except for radical reactants. Considering both barriers and the associated tunneling crossover temperatures, the most favorable hydrogenation sequence starts from HNC rather than from HCN. The network initially evolves toward either methanimine (\ce{H2CNH}; which emerges as the central species) or the singlet open-shell carbene \ce{^1[HC{:}NH2]}. Subsequent hydrogenation steps lead to methylamine (\ce{CH3NH2}). Along these reaction paths, several processes are barrier-less (e.g., the hydrogenation of molecular radicals), while some H-abstraction reactions can compete with addition channels. Reactions involving \ce{H2} as a reactant are rare, as most of these channels are endoergic. Deuterium substitution has only a minor impact on classical activation energies but significantly affects the transition-state imaginary frequencies and, therefore, the expected tunneling efficiencies.}
   {Our results support the idea that methanimine and methylamine can efficiently form from HNC on cold interstellar ices, with methanimine acting as a chemical sink within the network. HCN, on the other hand, is less reactive and hence has a higher chance of being preserved. The reaction network presented here provides quantitative constraints for astrochemical models of nitrogen-bearing organic chemistry in star-forming regions.}


   \maketitle

\section{Introduction}

Numerous CN-bearing species have been detected in the interstellar medium (ISM), ranging from simple molecules such as \ce{HCN} and \ce{CH3CN} to larger ones such as cyanopolyynes. Owing to their large dipole moments, these molecules are readily detectable in the gas phase using radioastronomical facilities, and as a result, they are routinely employed by astronomers to trace the physical conditions of the ISM, both in the Milky Way and in external galaxies \citep[e.g.,][]{Schilke1992, Hacar2020, LeGal2014, Oberg2015Natur, jin2015hcn, Ilee2021ApJS, Tennis2023MNRAS}.

From a chemical perspective, nitriles can appear as two different isomers, such as \ce{HCN} and \ce{HNC} or \ce{CH3CN} and \ce{CH3NC}. They are particularly interesting in the context of prebiotic chemistry. For example, they can hydrolyse into carboxylic acids under suitable conditions \citep[e.g.,][]{miller1953,miller1959}, a process that potentially links them to the formation of amino acids (as exemplified by the Strecker synthesis; \citealt{strecker_ueber_1854}), among other prebiotic species important for the RNA-world hypothesis \citep[see, e.g.,][]{oro1961amino,oro1961mechanism,powner2009synthesis,patel2015common,becker2019unified,cappelletti2026,JimenezSerra2020}.

The simplest closed-shell CN-bearing species in the ISM are HCN and HNC, which have been extensively studied in the literature. Their relative abundances are often used as a diagnostic tool for temperature \citep{Hacar2020}, and their emission can be used to trace dense gas regions; both are critical parameters for probing the ISM and other galaxies.

Both HCN and HNC can be formed through gas-phase processes or on the surfaces of interstellar icy grains \citep[see, e.g.,][]{enrique2024complex}. Therefore, subsequent chemical evolution on interstellar ices should lead to more complex species. Observational evidence points to HCN surface hydrogenation as the main path to methylamine formation \citep[e.g.,][]{Suzuki2023}; this is supported by the experiments of \citet{Theule2011}, which showed that HCN hydrogenation leads to \ce{CH3NH2}, with \ce{CH2NH} as a likely intermediate. With computational chemistry, \citet{Woon2002} investigated the hydrogenation of \ce{HCN} and \ce{CH2NH} in the context of interstellar amorphous solid water ices, using an implicit solvation model to account for the effect of the ice environment (i.e., without explicitly including water molecules). Their results show that hydrogenation at the carbon atom of HCN is significantly more favorable, with a lower activation barrier ($\sim$30.5~\kjmol\footnote{To convert \kjmol into K, one only needs to multiply by 120.27 K/\kjmol.}) compared to hydrogenation at the nitrogen atom ($\sim$53.5~\kjmol). A similar trend was observed for \ce{CH2NH}, with barriers of approximately 19.2~\kjmol (C-hydrogenation) and 25.5~\kjmol (N-hydrogenation). More recently, \citet{molpeceres2024carbon} investigated the reaction of \ce{^3C + NH3} on interstellar ice surfaces. As part of their reaction network, hydrogenation pathways of HCN and HNC were also considered, although limited to atomic hydrogen and excluding molecular hydrogen (\ce{H2}) reactions.

More broadly, numerous experimental studies have investigated the energetic processing of interstellar ice analogs containing cyanide-bearing molecules. Small nitriles such as \ce{HCN} and \ce{CH3CN} are highly reactive when exposed to UV irradiation \citep[e.g.,][]{Hudson2000,Gerakines2004,Bernstein2002,Danger2011VUV,JEscobar2014,Bulak2021} or X-rays \citep[e.g.,][]{Volosatova2021,Basalgete2021,Carvalho2022}, as well as under ion and electron bombardment \citep[e.g.,][]{Hudson2000,Gerakines2004,Hudson2008}. These energetic processes lead to the formation of a wide range of products, including amines, amides, aminonitriles, and salts such as \ce{OCN-}, particularly when ammonia is present. A recurring observation in these experiments is that the presence of water enhances chemical complexity, while both UV and X-ray irradiation can induce desorption processes. In addition, \ce{HCN} is known to undergo polymerization under energetic processing \citep[e.g.,][]{Gerakines2004,Volosatova2021}.

Despite the chemical richness observed in these experiments, the interpretation of results is often complicated by the lack of specificity in the reaction pathways. Multiple processes occur simultaneously, making it difficult to isolate individual reaction channels or derive quantitative kinetic data and branching ratios. One example of a different approach is the work of \citet{Carvalho2022}, who combined soft X-ray irradiation experiments on \ce{CH3CN} with a chemical network model to interpret the observed chemistry. Thanks to the use of a model and concomitant input parameters, they found that methyl cyanide can largely survive energetic processing. 

In this context, theoretical studies such as the present one can help disentangle individual reaction pathways and provide the energetic information required to guide the interpretation of laboratory experiments and astrochemical models. At the same time, dedicated experiments targeting specific hydrogenation pathways of \ce{HCN} and \ce{HNC} on water-rich ice surfaces would be highly valuable for testing the mechanisms proposed here.
To address these challenges and build upon our previous studies \citep{enrique2024complex,EnriqueRomero2025}, we revisited the surface destruction pathways of HCN and HNC via successive hydrogen addition and abstraction reactions involving \ce{H^.} and \ce{H2}.

We aim to connect these two molecules to their fully hydrogenated counterpart, methylamine (\ce{CH3NH2}).
This is in qualitative agreement with the reaction sequence proposed by the only laboratory work available in the literature, \citet{Theule2011}, who connected \ce{HCN} and \ce{HNC} to methylamine through methanimine.
Notice that the methanimine is also relevant in the prebiotic formation of glycine (\ce{NH2CH2COOH}) via the Strecker synthesis, where methanimine reacts with HCN \citep[e.g.,][]{Danger2011_aminoacetonitrile}.

\section{Methods} \label{sec:methods}

\begin{figure}[!htbp]
    \centering
    \includegraphics[width=0.5\textwidth]{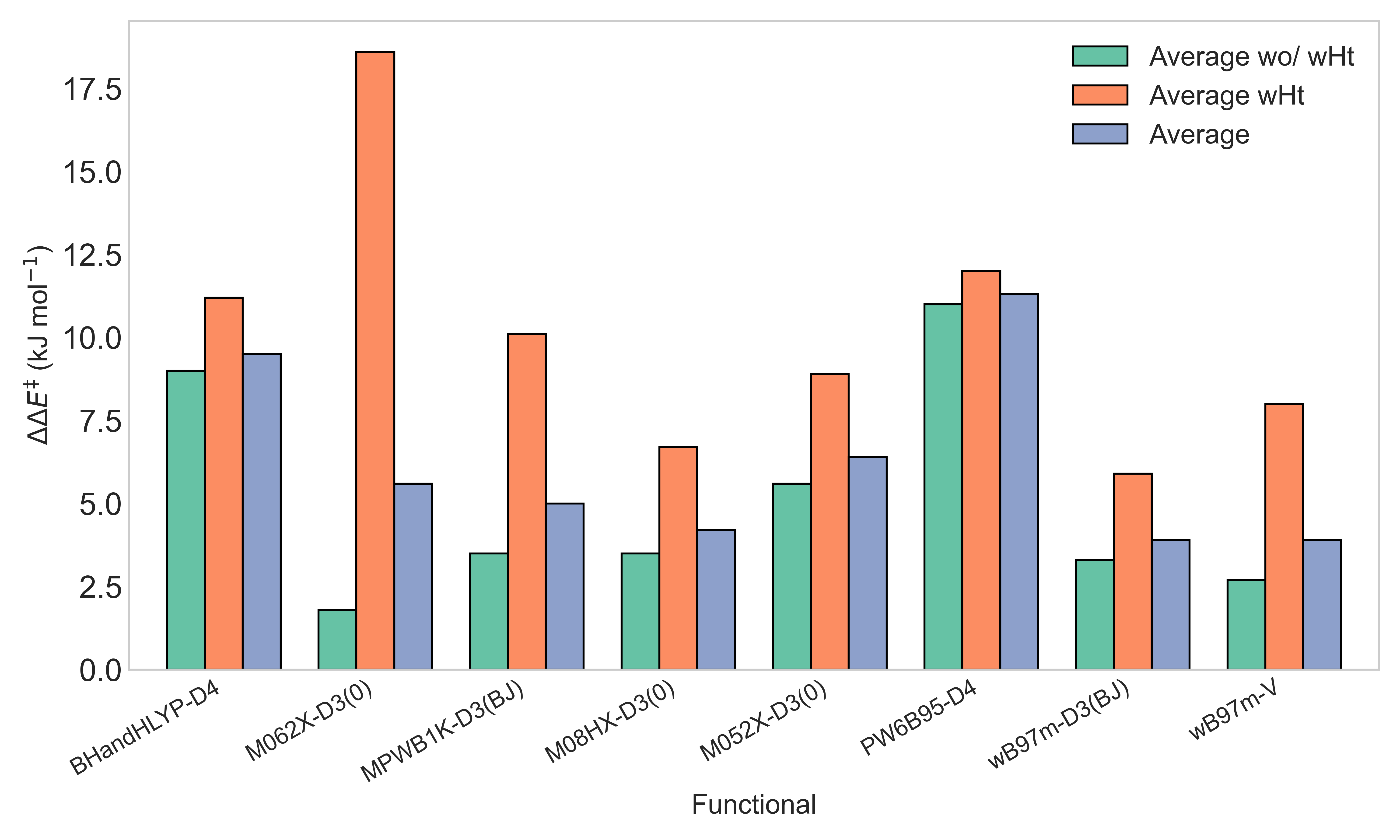}
    \caption{Selected functionals from the benchmark study. Three metrics are reported: (i) the average energy deviation excluding proton-transfer reactions, (ii) the average energy deviation for proton-transfer reactions only, and (iii) the global average of all reactions. These values are obtained from 13 benchmarked reactions (see the main text for details and the appendix for tabulated data).}
    \label{fig:benchmark}
\end{figure}

All calculations were performed using {\sc Orca} 6.0.1 \citep{neese2020orca, Neese2022}. Benchmarked density functional calculations were used across this work. Given the number of reactions studied in this work, we chose the triple-$\zeta$-quality ma-def2-TZVP basis set \citep{zheng2011minimally}, balancing computational cost and accuracy for density functional calculations. We used multiple computational algorithms for this work, encompass geometry optimizations, transition state searches, intrinsic reaction coordinate calculations to connect stationary points along the reaction pathway on the potential energy surface, potential energy surface scans, nudged elastic band calculations \citep{asgeirsson2021nudged}, and frequency calculations to ensure that our geometries correspond to minima or first-order saddle points for transition states. Tight self-consistent field (\texttt{TightSCF}) convergence criteria and dense integration grids (\texttt{DEFGRID3}).
Unrestricted Kohn-Sham orbitals were used consistently, and for all open-shell singlet calculations, we employed the broken-spin-symmetry approach via {\sc Orca}'s spin-flip routine.
A comprehensive benchmark comprising 13 reactions was performed \citep[raw data available on Zenodo][]{zenodo_dataset}, where the performance of a wide range of functionals in combination with the ma-def2-TZVP basis set was tested against CCSD(T)/aug-cc-pVTZ reference data. Figure~\ref{fig:benchmark} summarizes the results by reporting average energy deviations for selected functionals. Some functionals exhibit a marked difference in performance between \ce{H}/\ce{H2} addition reactions and water-assisted H-transfer (wHt) reactions. To make this distinction clear, we report the results for these two subsets separately in Fig. \ref{fig:benchmark}, in addition to providing the overall average for reference.
For the wHt reactions, $\omega$B97m-D3(BJ) yields the highest accuracy. In contrast, for \ce{H}/\ce{H2} additions and abstraction reactions, M06-2X-D3(0) performs best. We therefore adopted a combined strategy, applying each functional to the reaction class for which it is most reliable. In addition, the use of M06-2X-D3(0) is further supported by its successful application in our previous studies, thereby ensuring methodological consistency. From the reactions listed in Table ~\ref{Table:summary_and_compare}, we took H-addition reactions \ref{chem:HCN+H__H2CN}, \ref{chem:HCN+H__cis/trans-HCNH}, \ref{chem:HNC+H__CNH2}, \ref{chem:HNC+H__cis/trans-HCNH}, and \ref{chem:H2CNH+H__H2CNH2}; the H-abstraction reactions \ref{chem:H3CNH+H__tripCH3N+H2}, \ref{chem:CH3NH2+H__H2CNH2+H2} and \ref{chem:CH3NH2+H__H3CNH+H2}; \ce{H2}-hydrogenation reactions \ref{chem:H2CNH+H__cis/trans-HCNH+H2} from cis- and from trans-\ce{HCNH}; and the wHt reactions \ref{chem:CNH2__cis/trans-HCNH}, \ref{chem:HCNH2__H2CNH}, and \ref{chem:H3CNH__H2CNH2}.

Finally, M06-2X-D3(0) performs poorly for the H-abstraction reactions \ref{chem:singHCNH2+H__CNH2+H2} and \ref{chem:HCNH3+H__CNH3+H2}, yielding unrealistic ZPE corrections, exceeding 50~\kjmol. For these cases, we instead employed the next-best functional from the benchmark, $\omega$B97m-D3(BJ), which produces physically reasonable ZPE corrections, with corrected activation barriers of 8.8 and 6.6~\kjmol for reactions~\ref{chem:singHCNH2+H__CNH2+H2} and \ref{chem:HCNH3+H__CNH3+H2}. In agreement with the literature \citep{molpeceres2024carbon}.

\section{Results}\label{sec:results}

\subsection{Chemical network}\label{sec:Network}

\begin{figure*}[!htbp]
    \centering
    \includegraphics[width=\textwidth]{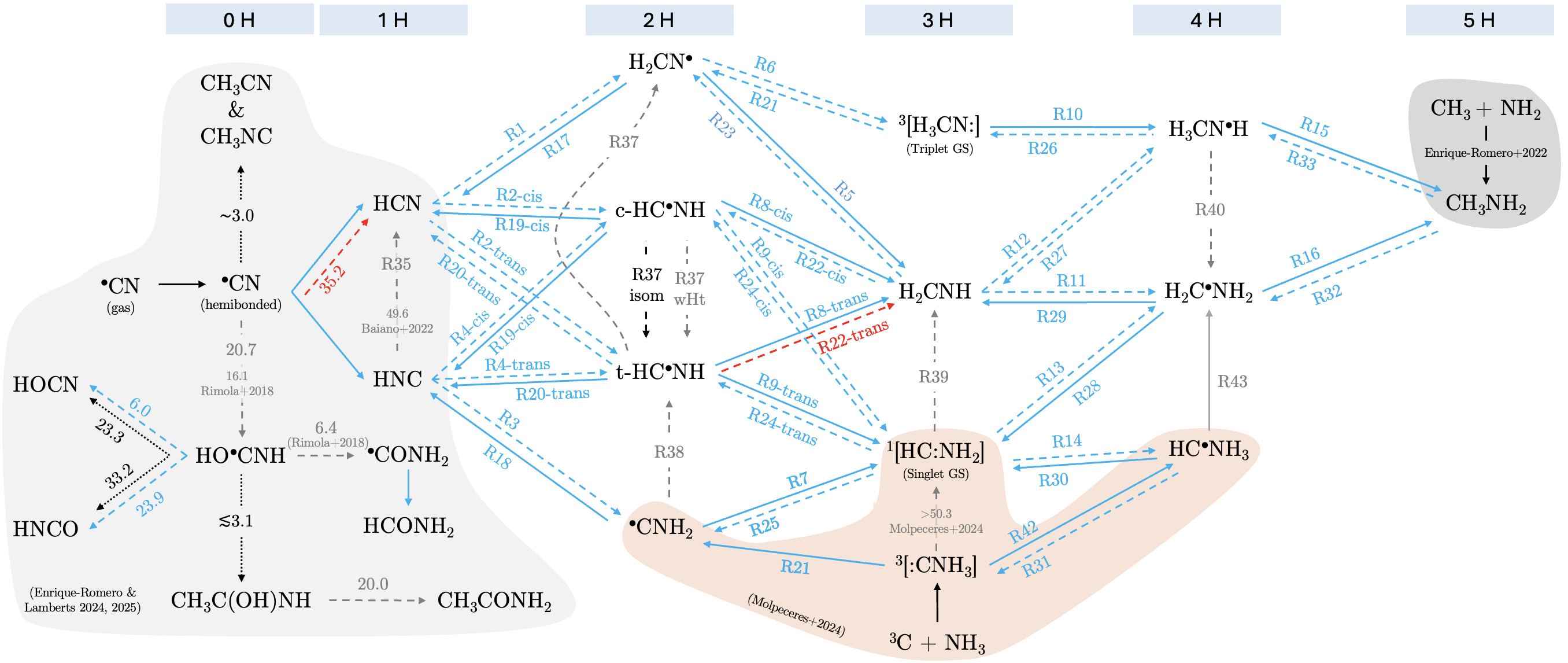}
    \caption{Schematic representation of the \ce{^.CN} chemical network on interstellar ice surfaces. The figure summarizes the reactions investigated in this work (the hydrogenation of HCN and HNC) together with related processes reported in previous studies \citep{ER2022ApJS,molpeceres2024carbon,enrique2024complex,ER2025_CH3CN}, shown as colored regions. Unless otherwise indicated, radical species have doublet spin multiplicity. Solid arrows denote barrier-less reactions, whereas dashed arrows indicate reactions with an activation barrier. Activation energies are given in \kjmol. Arrow colors indicate reaction types: blue for H addition and abstraction, red for reactions with \ce{H2}, gray for wHts, and black for other processes (e.g., isomerization or radical–radical reactions). Energies are reported on a per-reaction basis, assuming efficient thermalization on amorphous solid water, such that each activation barrier is referenced to the products of the preceding step.}
    \label{fig:summary}
\end{figure*}

Four main types of reactions are considered: (i) H-additions involving atomic hydrogen, (ii) H-additions involving molecular hydrogen, (iii) H-abstractions by H atoms from molecular species (both closed-shell and radicals), and (iv) wHt isomerization reactions. To facilitate the presentation of our results, we organized the reactions by the number of H atoms added to HCN and HNC up to the five hydrogen atoms of \ce{CH3NH2}, and provided the chemical network in Fig.~\ref{fig:summary}. Furthermore, Table~\ref{Table:summary_and_compare}, provides the complete list of reactions with complementary energetic data and a comparison with works in the literature. To aid the reader, we have also included a condensed version of the full reaction network in Fig. \ref{fig:concl_summary}, which details the reaction pathways that we deem most important for astronomical purposes.

\subsection{One hydrogen addition}\label{sec:first-hydrogenation}

\begin{figure*}[!htbp]
    \centering
    \includegraphics[width=0.7\linewidth]{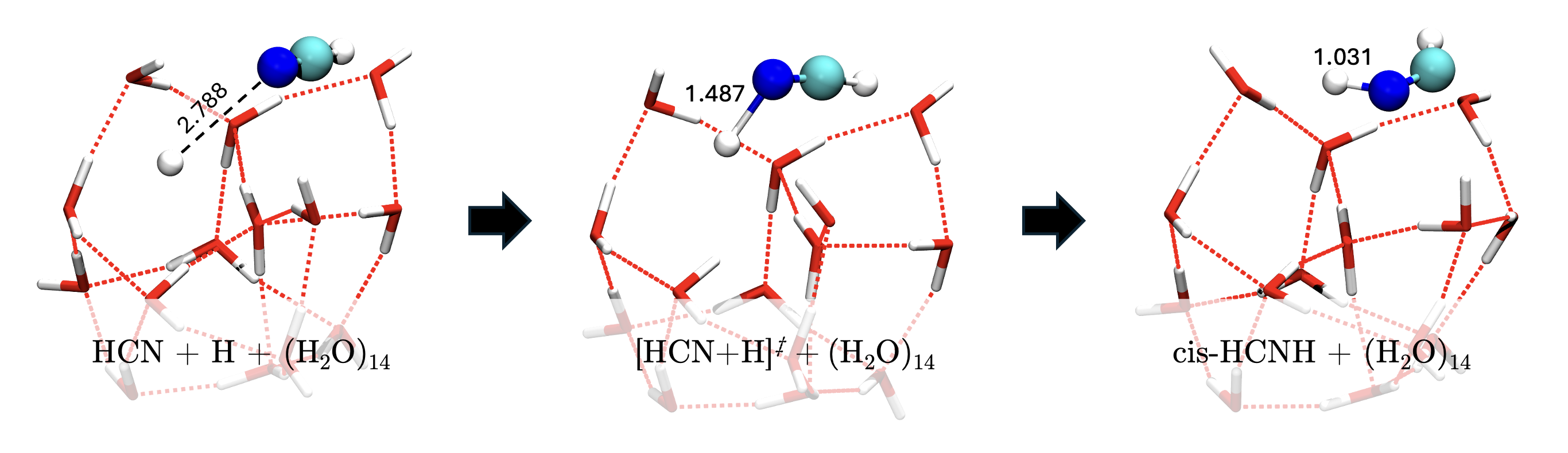}
    \includegraphics[width=0.7\linewidth]{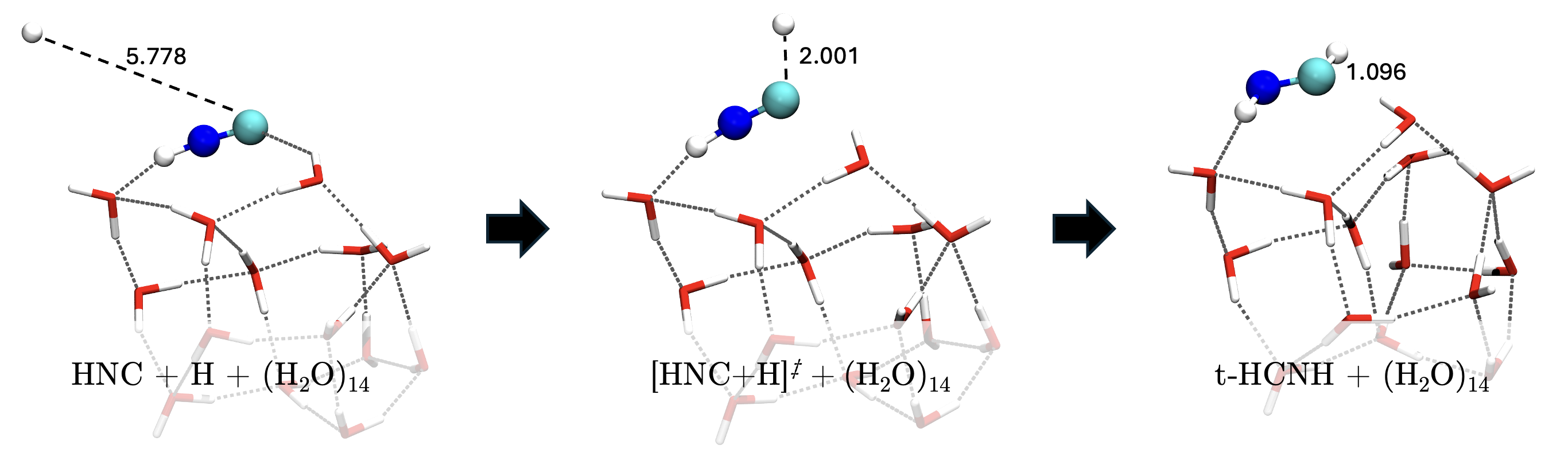}
    \caption{Geometries of the H-addition reactions on HNC leading to cis- and trans-\ce{HC^.NH} (reactions \ref{chem:HNC+H__cis/trans-HCNH}-cis and \ref{chem:HNC+H__cis/trans-HCNH}-trans). Distances are in \AA.}
    \label{fig:HNC_evo}
\end{figure*}

Six reactions convert HCN and HNC to species with the \ce{\{H2CN\}} general formula. The lowest activation energy barriers correspond to \ce{HNC} hydrogenation, forming either the cis- or trans-isomers of \ce{HC^.NH} (reaction~\ref{chem:HNC+H__cis/trans-HCNH}, Fig. \ref{fig:HNC_evo}), with barriers of 8.2 and 12.5~\kjmol, respectively, while the barrier to form \ce{C^.NH2} is 55.9 \kjmol. In contrast, H-addition to the more stable isomer \ce{HCN}  (reactions~\ref{chem:HCN+H__H2CN} and~\ref{chem:HCN+H__cis/trans-HCNH}), involves substantially higher barriers of approximately 28.4 to form \ce{H2CN^.}, 33.9~\kjmol to form the cis-\ce{HC^.NH} and 41.7 for the trans-\ce{HC^.NH} isomer.
The \ce{HNC \to HCN} conversion is disfavoured, as we find water-assisted proton transfer reactions (reaction~\ref{chem:HCN__HNC}-wHt) exhibit large activation barriers, at least 35.1~\kjmol when mediated by three water molecules, in agreement with the work by \citet{baiano2022gliding}, where a barrier of 49.6~\kjmol for the same process involving two water molecules is reported.

Given the relatively high barriers, all these reactions are expected to proceed primarily via quantum tunneling given the relatively high crossover temperatures (see Table~\ref{Table:summary_and_compare}), with the lowest barrier for reaction \ref{chem:HNC+H__cis/trans-HCNH}-cis (\ce{HNC} to \ce{cis-HC^.NH}). Nonetheless, these first-step hydrogenation products are likely to be short-lived intermediates, as further H-addition or abstraction reactions can take place, with either small or no barriers at all.

In summary, the more reactive isomer between HCN and HNC toward H-addition is HNC, which leads to \ce{HC^.NH}.

\subsection{Two hydrogen additions}

Continuing along the hydrogenation sequence toward the \ce{\{H3CN\}} family, the intermediates produced from the first H-addition (cis- and trans-\ce{^.HCNH}, \ce{C^.NH2}, and \ce{H2CN^.}), undergo further hydrogenation. Within the \ce{\{H3CN\}} family, the most stable isomer is methanimine (\ce{H2C=NH}), already hypothesized to be an intermediate along the reaction path by \citet{Theule2011}.

The H-addition reactions on \ce{\{H2CN\}} leading to methanimine are barrier-less (reactions~\ref{chem:cis/trans-HNCH+H__H2CNH}, \ref{chem:H2CN+H__H2CNH}). In contrast, the formation of aminocarbene and methyl nitrene involves activation barriers of 25.8 and 13.5~\kjmol, corresponding to reactions~\ref{chem:cis/trans-HNCH+H__singHCNH2} (from the cis isomer) and~\ref{chem:H2CN+H__tripH3CN}, respectively. The remaining pathways, those involving the trans-\ce{^.HCNH} isomer and the reaction of \ce{C^.NH2} with \ce{H^.} (reactions~\ref{chem:cis/trans-HNCH+H__singHCNH2} and~\ref{chem:CNH2+H__singHCNH2}), are also barrier-less. The reactivity with molecular hydrogen (\ce{H2}) is limited to \ce{t-HC^.NH + H2 \to H2CNH + H^.} (reaction~\ref{chem:H2CNH+H__cis/trans-HCNH+H2}), which exhibits a barrier of 36.6~\kjmol. Turning now toward H-abstraction reactions, the destruction of all \ce{H2CN} isomers back to \ce{HCN} and \ce{HNC} via H-abstraction with an H atom proceeds without activation barriers.

\begin{figure*}[!htpb]
    \centering
    \includegraphics[width=0.7\linewidth]{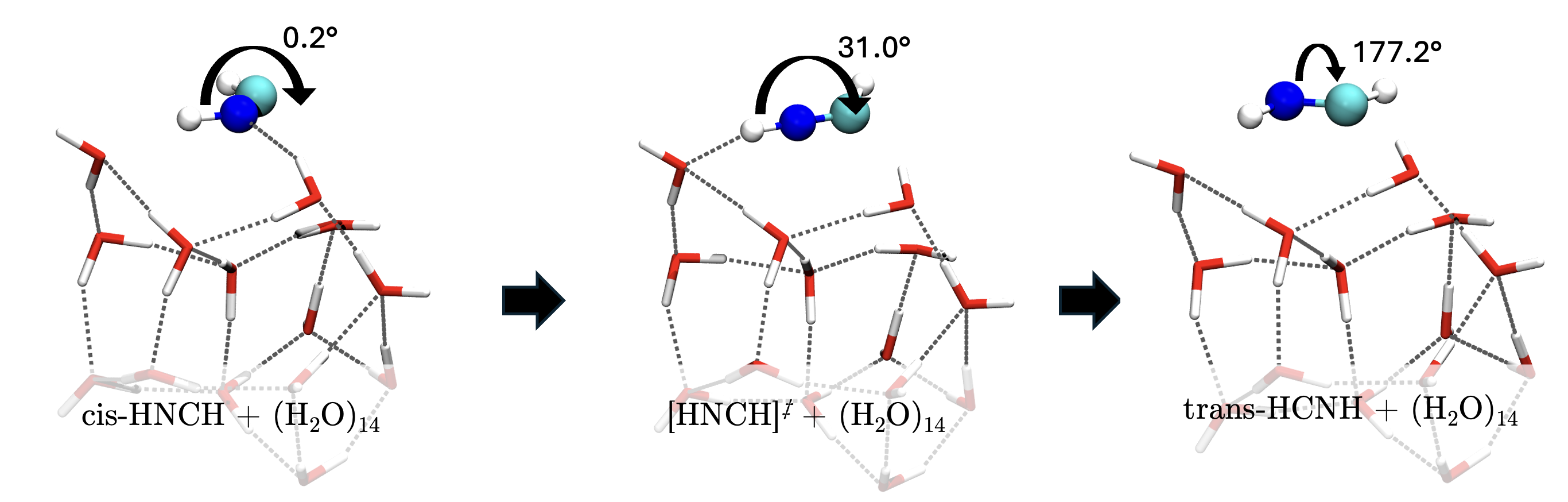}
    \includegraphics[width=0.7\linewidth]{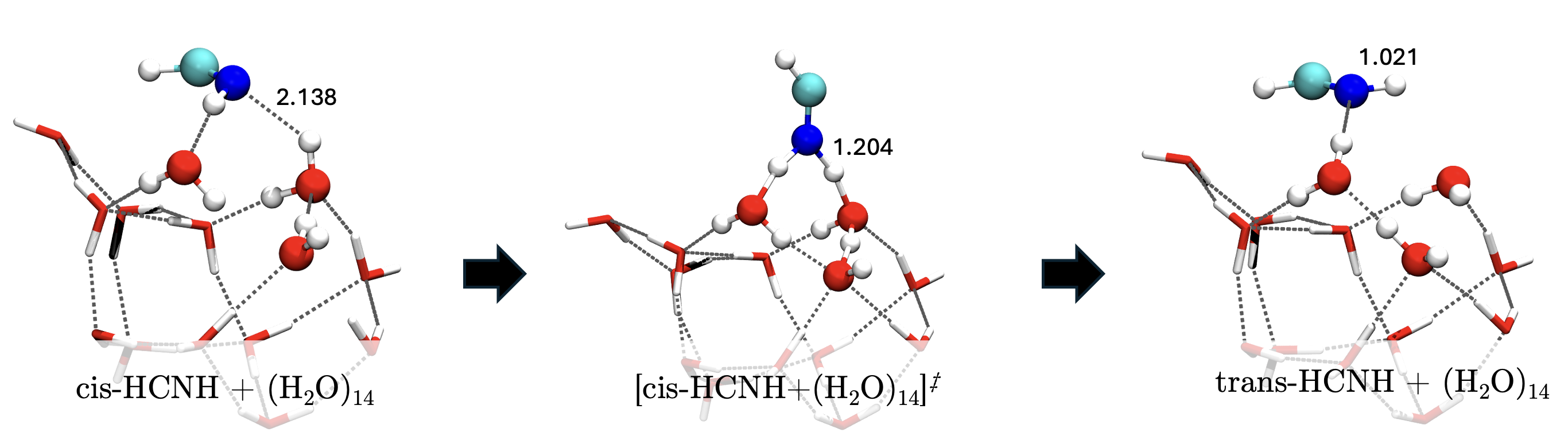}
    \caption{Geometries of the isomerization reactions from cis- to trans-\ce{HC^.NH} (reactions \ref{chem:cis-HCNH__trans-HCNH}-isom and \ref{chem:cis-HCNH__trans-HCNH}-wHt). In the upper panel, the reaction coordinate is the torsion angle of \ce{HC^.NH}. Distances are in \AA.}
    \label{fig:cis-trans_isom}
\end{figure*}

The interconversion between cis- and trans-\ce{HCNH} can proceed either via direct isomerization (torsion around the dihedral angle) or through water-assisted proton transfer (reaction~\ref{chem:cis-HCNH__trans-HCNH}). The direct pathway involves a barrier of 23.9~\kjmol, whereas the water-mediated mechanism exhibits a significantly higher barrier of 65.5~\kjmol over three water molecules (see Fig.~\ref{fig:cis-trans_isom}). In addition, two further wHt isomerization reactions were identified, namely \ce{trans-HC^.NH ->[wHt][] H2CN^.} and \ce{C^.NH2 ->[wHt][] trans-HC^.NH}, with activation barriers of 83.7 and 30.8~\kjmol, respectively.

Finally, it must be noted that methyl nitrene (\ce{^3[H3CN{:}]}) and aminocarbene (\ce{^1[HC{:}NH2]}) are higher-energy isomers of methanimine. These species exhibit non-octet electronic configurations at the nitrogen and carbon atoms, respectively, resulting in triplet and singlet ground states. To avoid ambiguity, we have explicitly indicated both their electronic state and the localization of the non-bonding electrons. Such spin configurations are characteristic of nitrenes and carbenes, where the distribution of unpaired electrons governs their reactivity and electronic structure. In the case of aminocarbene, the singlet ground state is stabilized by interaction between the lone pair of the amino group and the empty \textit{p} orbital of the carbene center, which lowers its energy relative to the corresponding triplet state.

In summary, the most probable process at this stage of hydrogenation is H-addition to trans \ce{HC^.NH}, leading preferentially to methanimine or aminocarbene formation. In contrast, the production of methyl nitrene is less likely, as it requires the hydrogenation of \ce{H2CN^.}, which itself forms via a barrier-mediated hydrogenation of \ce{HCN}.

\subsection{Three hydrogen additions}

\begin{figure*}[!htbp]
    \centering
    \includegraphics[width=0.7\linewidth]{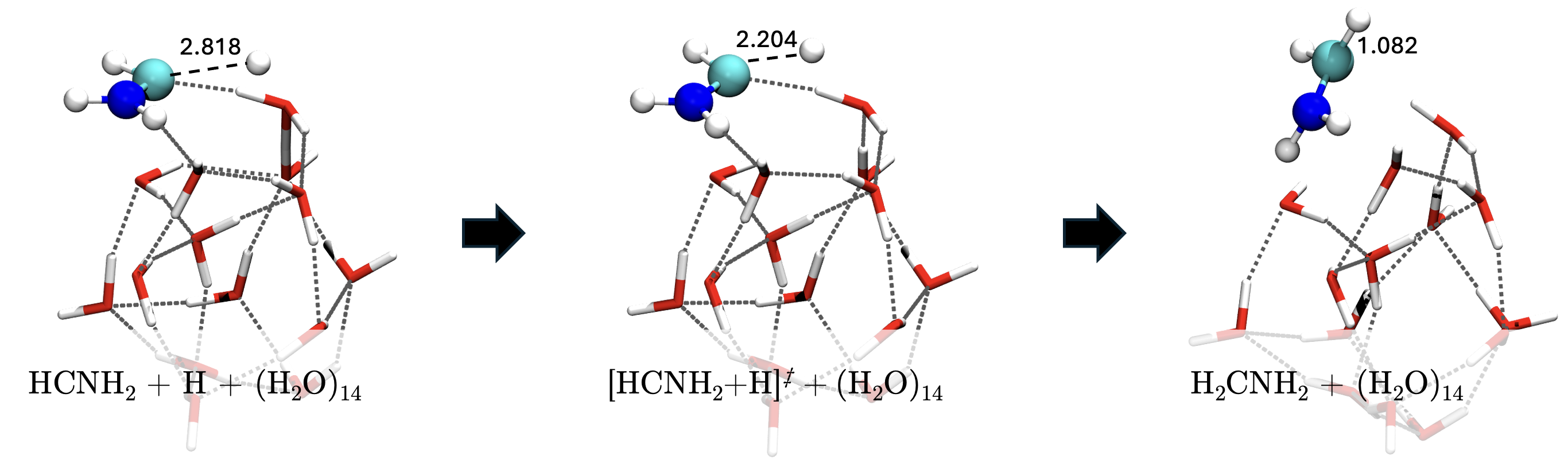}
    \includegraphics[width=0.7\linewidth]{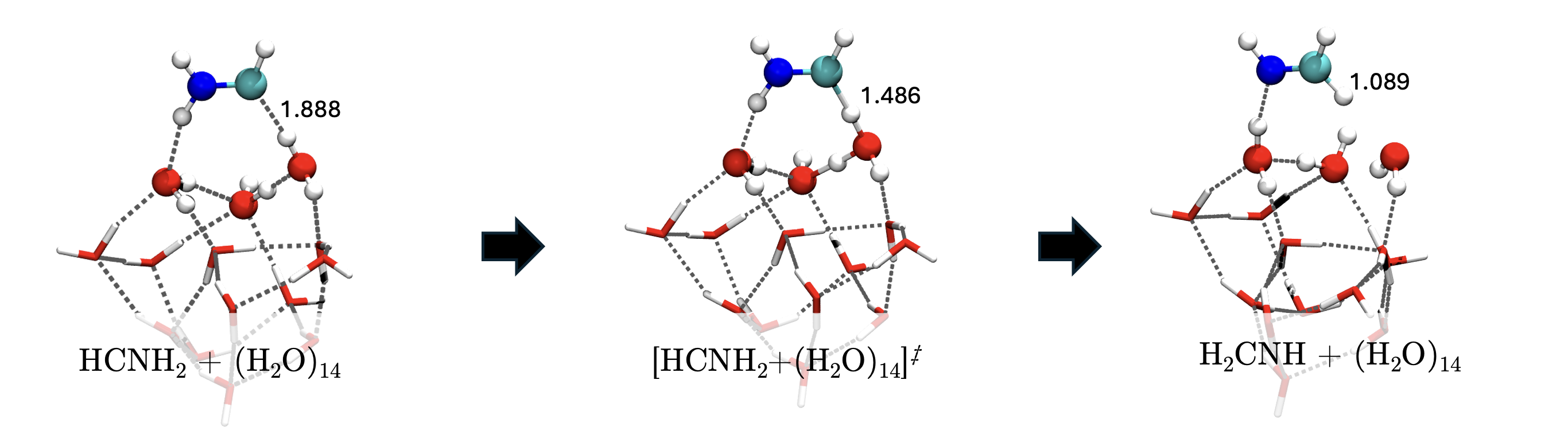}
    \caption{Molecular geometries of the lowest barrier channels involving the evolution of \ce{^1[HC{:}NH2]} into \ce{H2C^.NH2} (upper panels) or its water-assisted isomerization into \ce{H2CNH} (lower panels; reactions \ref{chem:singHCNH2+H__H2CNH2} and \ref{chem:HCNH2__H2CNH}). Distances are in \AA.}
    \label{fig:HCNH2evo}
\end{figure*}

From the \ce{\{H3CN\}} family, the main reactants involved in this third H-addition step are methanimine and aminocarbene. The hydrogenation of the former into \ce{H3CN^.H} or \ce{H2C^.NH2} (reactions~\ref{chem:H2CNH+H__H3CNH} and~\ref{chem:H2CNH+H__H2CNH2}) exhibits activation barriers of 16.0 and 22.9~\kjmol, respectively. In contrast, the conversion of aminocarbene to \ce{H2C^.NH2} (reaction~\ref{chem:singHCNH2+H__H2CNH2}, Fig. \ref{fig:HCNH2evo}) proceeds with a very low barrier of only 4.0~\kjmol, making this the most favorable pathway.
On the other hand, hydrogenation of the less relevant \ce{^{3}[H3CN{:}]} isomer is barrier-less, leading directly to \ce{H3CN^.H} (reaction~\ref{chem:tripH3CN+H__H3CNH}).

In addition, the isomerization through wHt from aminocarbene to methanimine sports a very low barrier, of just 4.9 \kjmol over three water molecules, indicating that methanimine might be a chemical ``sink'' in the sense that it is a main hub in the chemical network. Finally, we found \ce{HC^.NH3} to be highly unstable, as it barrier-lessy evolves into \ce{H2C^.NH2} through a wHt reaction or can come back to \ce{^3[{:}CNH3]}.

Hydrogen abstraction reactions from the \ce{\{H3CN\}} family are generally characterized by the presence of activation barriers. Methanimine is relatively resistant to destruction via this mechanism, as the only exothermic pathway leads to \ce{cis-HC^.NH} over a barrier of 24.8~\kjmol (reaction \ref{chem:H2CNH+H__cis/trans-HCNH+H2}). Aminocarbene can be converted back to \ce{C^.NH2} (reaction~\ref{chem:singHCNH2+H__CNH2+H2}), over a barrier of 8.8~\kjmol or to \ce{trans-HC^.NH} (reaction~\ref{chem:singHCNH2+H__cis/trans-HCNH+H2}) with a barrier of 14.8 \kjmol. On the other hand, methyl nitrene is more prone to being destroyed into \ce{H2CN^.} with a significantly lower barrier of only 6.7~\kjmol (reaction \ref{chem:tripH3CN+H__H2CN+H2}).

In summary, forming either methanimine or aminocarbene ensures the continuation of the hydrogenation sequence rather than cycling back to the \ce{\{H2CN\}} family. In contrast, the route involving \ce{^{3}[H3CN{:}]} is of limited relevance, as its formation is kinetically unlikely and it is readily destroyed once formed.

\subsection{Four hydrogen additions}

The key radical intermediates \ce{H3CN^.H} and \ce{H2C^.NH2}  both undergo barrier-less reactions with atomic hydrogen to produce the fully hydrogenated product, methylamine (\ce{CH3NH2}). These radicals are accessible after the hydrogenation of methylamine and aminocarbene; the latter is the faster process, given its low energy barrier.
The H-abstraction reaction leading \ce{H2C^.NH2} back to methanimine (reaction~\ref{chem:H2CNH2+H__H2CNH+H2}) is barrier-less, while the destruction of \ce{H3CN^.H} back to either methanimine or \ce{^{3}[H3CN{:}]} are characterized by barriers of 4.8 and 15.6 \kjmol. We tried to also simulate the destruction of \ce{H2CNH2} into \ce{^1[HC{:}NH2]}, all our attempts ended up in the H-addition product, methylamine. These results further support methanimine's role as a sink in the network. The isomerization of \ce{H3CN^.H} into \ce{H2C^.NH2} (reaction~\ref{chem:H3CNH__H2CNH2}) requires a high activation energy barrier of 69.3 \kjmol over two water molecules.

Finally, once methylamine is formed, its destruction via H-abstraction to \ce{H3CN^.H} or \ce{H2C^.NH2} (reactions~\ref{chem:CH3NH2+H__H3CNH+H2} and \ref{chem:CH3NH2+H__H2CNH2+H2}) exhibit high barriers of 48.0 and 30.2 \kjmol, respectively. For the latter case, the nature of this barrier is related to the strong binding energy of the \ce{-NH2} group to the water ice, since, if the H atom is found right under this group, reaction~\ref{chem:CH3NH2+H__H2CNH2+H2} has a much lower activation energy barrier of just 9.9 \kjmol.

These results suggest that, for this final round of H-additions and abstractions, the network is strongly biased toward the accumulation of methylamine, as reverse pathways are generally hindered by activation barriers and local ice effects.

\subsection{Deuteration}\label{sec:deuter}

\begin{figure}[!htbp]
    \centering
    \includegraphics[width=0.45\textwidth]{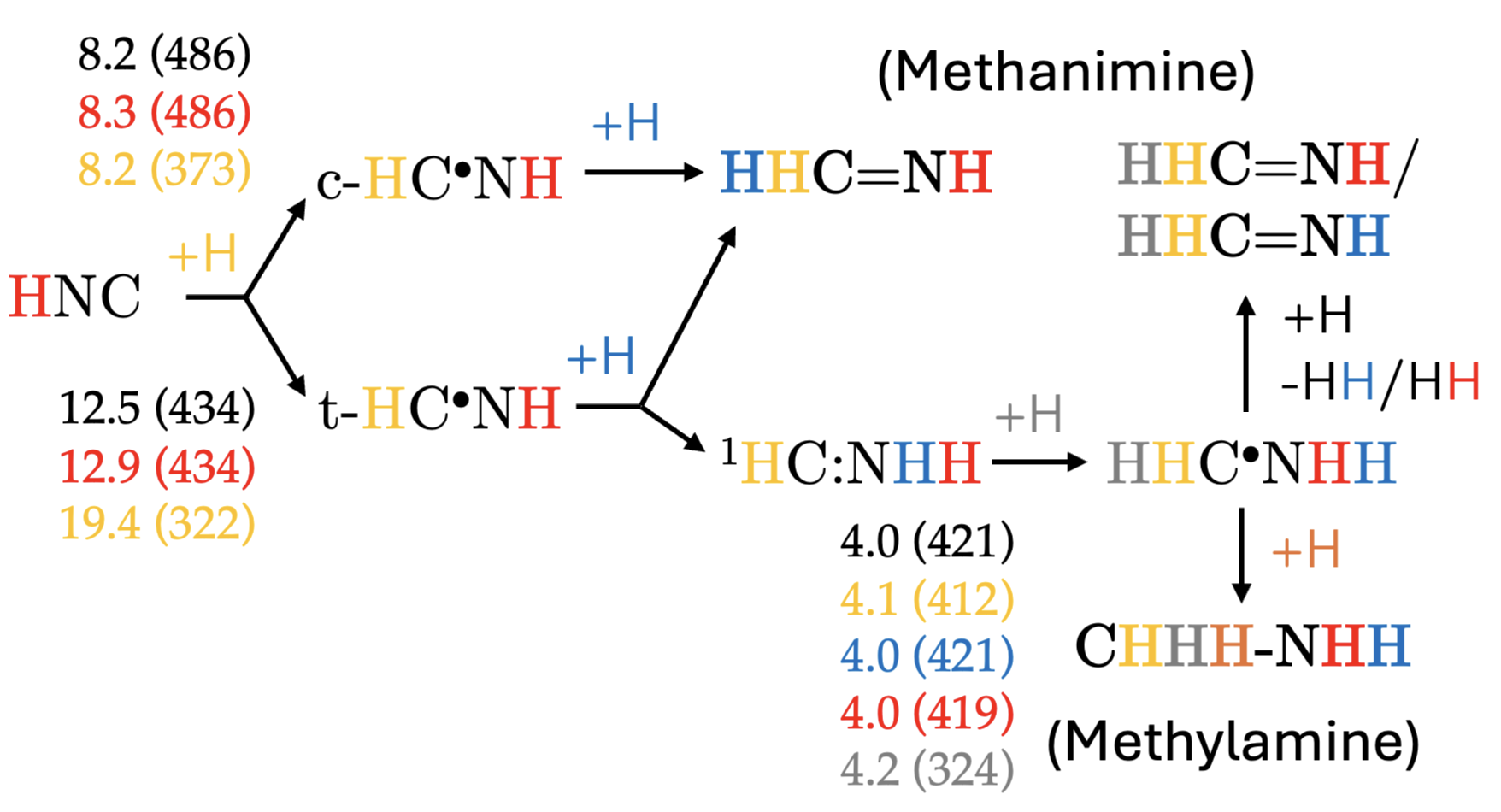}
    \caption{Key D-substitution patterns for the energetically favored pathways leading to methanimine and methylamine from HNC and \ce{^1[HC{:}NH2]}. Single deuterium substitutions are shown as color-coded H atoms. Bare values correspond to ZPE-corrected activation energies (in~\kjmol; when present), and values in parentheses indicate the transition-state imaginary frequencies ($|\nu^{\ddagger}|$; in cm$^{-1}$). Reaction steps without any activation energy value are barrier-less.}
    \label{fig:deuteration_Hadd}
\end{figure}

Deuterium substitution was investigated for a selected set of kinetically relevant reactions involving HNC and \ce{^1[HC{:}NH2]}, i.e., the energetically most favorable paths that contribute to the evolution toward methanimine and methylamine. Figure~\ref{fig:deuteration_Hadd} summarizes the possible products arising from D-substitution for H-addition reactions. A clear kinetic isotope effect is observed only when a D atom is added, which manifests primarily in the imaginary frequencies of the transition states, with shifts on the order of about -100~cm$^{-1}$, while the corresponding changes in activation energies are generally small. An exception is the reaction \ce{HNC + H \to t-HC^.NH}, which already presents a relatively high activation barrier compared to the formation of the cis isomer and is, therefore, inherently less efficient.

A closer inspection of Fig.~\ref{fig:deuteration_Hadd}, taking into account both the energetics and the changes in barrier widths, suggests that the direct addition of a D atom to either HNC or \ce{^1[HC{:}NH2]} is less efficient than the addition of an H atom to a parent molecule that already contains a \ce{D^.} substitution, under the assumption that tunneling is mandatory. From the color-coded single D-for-H substitutions, it appears that the reaction \ce{DNC + H^.} preferentially leads to \ce{H2CND}, whereas \ce{HDCNH} would more likely form via \ce{c-HC^.NH + D^.}. Given the relatively high barrier associated with the formation of the trans-\ce{HC^.NH} isomer, the production of \ce{^1[HC{:}NH2]} is more likely to proceed through the reaction \ce{^3C + NH3} \cite{molpeceres2024carbon}. In this case, and following our previous discussion of H-addition and abstraction reactions involving \ce{H2C^.NH2}, it is energetically more favorable to add an H atom to the carbon center rather than abstract one of its hydrogen atoms, while abstraction from the nitrogen moiety remains feasible. Consequently, any D-substitution pattern in methylamine can be produced, whereas for methanimine, assuming the deuterium originates from \ce{NH2D}, the formation of \ce{H2CND} would be favored. Alternatively, if the D-addition proceeds via \ce{H2C^.NH2 + D}, the preferential product would be \ce{HDCNH}.

\begin{figure}[!htbp]
    \centering
    \includegraphics[width=0.5\textwidth]{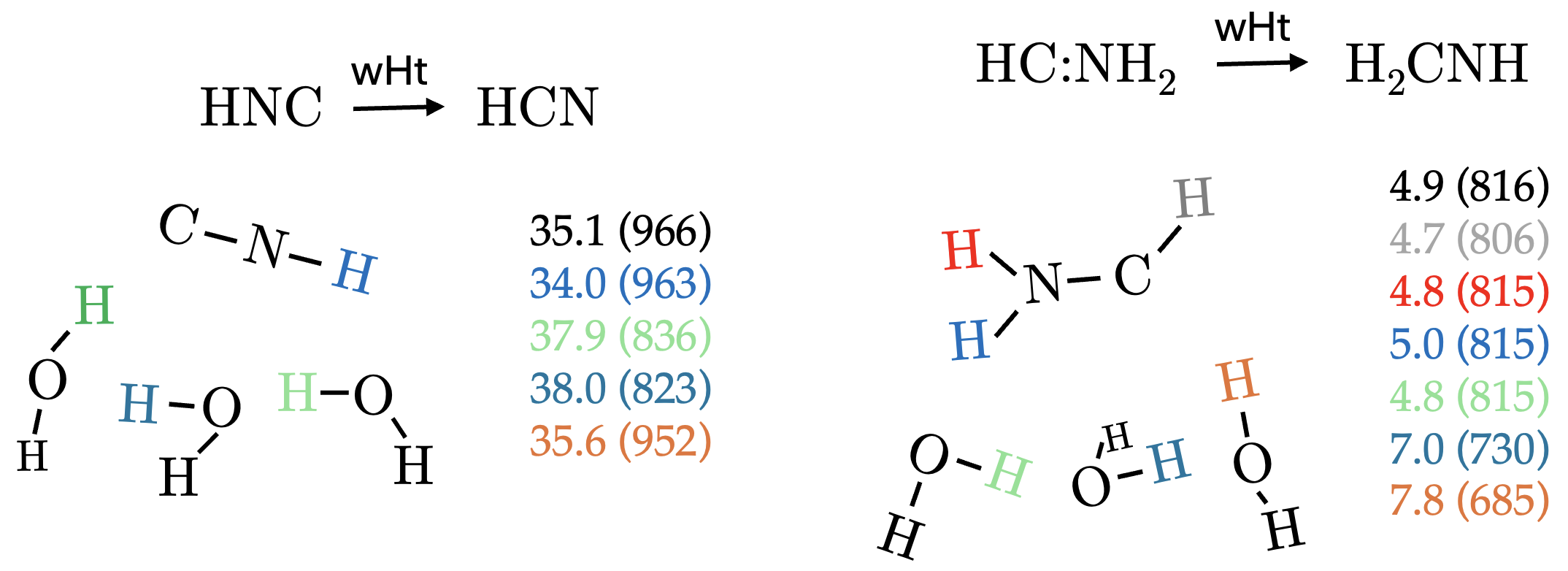}
    \caption{Key D-substitution patterns for the wHt-mediated isomerization of HNC to HCN and \ce{^1[HC{:}NH2]} to \ce{H2CNH} over three water molecules. Single deuterium substitutions are indicated by color-coded H atoms. Bare values give ZPE-corrected activation energies (in~\kjmol; when present), while values in parentheses denote the transition-state imaginary frequencies ($|\nu^{\ddagger}|$; in cm$^{-1}$).}
    \label{fig:deuteration_wHt}
\end{figure}

Among the isomerization channels, only the wHt reaction \ce{^1[HC{:}NH2] -> H2CNH} presents a low activation barrier (4.9~\kjmol). As shown in Fig.~\ref{fig:deuteration_wHt}, D substitution on \ce{^1[HC{:}NH2]} has a negligible impact on both the activation energy and the transition-state imaginary frequency. However, D substitution on the water molecules participating in the wHt mechanism increases the barrier by approximately 2–3~\kjmol and reduces the imaginary frequency by about 100–130~cm$^{-1}$, indicating a higher and wider barrier.

For the wHt isomerization of HNC into HCN, a similarly weak dependence on D-substitution is observed when deuterium is initially located on DNC, although the absolute activation barriers remain comparatively high.

\section{Discussion}\label{sec:discussion}
\subsection{Comparing to other works in the literature}

Other works in the literature have partially explored the chemical network presented above, and here we provide a comparison between our results and previous ones, summarized in Table \ref{Table:summary_and_compare}. Notably, \citet{molpeceres2024carbon} examined several of the reaction paths presented here while investigating the condensation of carbon atoms on ammonia molecules on interstellar ices, focusing mainly on H-addition and H-abstraction reactions together with a limited number of wHt processes. While most reactions exhibit similar energetic trends in the two studies, some differences arise. These discrepancies can largely be attributed to three factors: (i) the local binding configuration on the ice surface, (ii) the influence of zero-point energy (ZPE) corrections, and (iii) methodological differences between the computational approaches.

Surface reactions on amorphous solid water depend strongly on the local binding environment, including variations in binding energies and in the hydrogen-bond network. On the one hand, for reaction~\ref{chem:CH3NH2+H__H2CNH2+H2}, \ce{CH3NH2 + H^. \to H2C^.NH2 + H2}, we identified two barriers depending on the position of the incoming H atom relative to the \ce{-NH2} group: 30.2 or 9.9~\kjmol (see Fig. \ref{fig:CH3NH2+H}). The higher value agrees well with literature values of 37.3 \citep{molpeceres2024carbon} and 39.7~\kjmol \citep{joshi2022chemical}. The lower barrier arises from a configuration where the H atom reduces the interaction of the \ce{-NH2} moiety with the ice surface, destabilizing the reactant complex and lowering the apparent barrier, also evident from the reaction energies (see Table \ref{Table:summary_and_compare}). While this configuration is likely unstable and transient in a dynamic system (the H atom could diffuse away), it highlights the strong influence of the local ice environment on the resulting energetics. On the other hand, the wHt isomerization between \ce{^1[HC{:}NH2]} and methanimine (reaction~\ref{chem:HCNH2__H2CNH}) illustrates the role of ZPE corrections. In our work, the activation barrier is only 4.9~\kjmol, whereas \citet{molpeceres2024carbon} and \citet{Ferrero2024} reported values of 22.7 and 14.0~\kjmol, respectively. In our calculations, the uncorrected $\omega$B97m-D3(BJ) barrier is $\sim$17~\kjmol, indicating that the ZPE correction plays a substantial role, while differences in the local hydrogen-bond environment may further contribute.

A similar case is reaction~\ref{chem:CNH2__cis/trans-HCNH}, for which \citet{molpeceres2024carbon} reported a barrier of 88.8~\kjmol for formation of the trans isomer, whereas we obtain a barrier of 30.8~\kjmol, much closer to their gas-phase value of 39.5~\kjmol. The high barrier reported in that study led the authors to focus their discussion on \ce{cis-HC^.NH}, excluding the trans isomer. Although our value is significantly lower, this reaction is still unlikely to occur under pre-stellar conditions due to its very low tunneling crossover temperature (52~K), supporting their overall conclusions.

\citet{molpeceres2024carbon} were also unable to identify suitable wHt channels for the conversion of \ce{cis-HC^.NH} into \ce{C^.NH2} and \ce{^.CH2N}. In their work, this limitation was attributed to the binding modes obtained for the reactants. From our perspective, however, it is more likely related to the greater geometric accessibility for H-transfer from the trans isomer compared with the cis configuration, which allows the wHt mechanism to operate more effectively.

A further discrepancy arises for reaction~\ref{chem:H2CNH2+H__H2CNH+H2}, which \citet{molpeceres2024carbon} considered non-viable on their ice model. In contrast, we find that this reaction proceeds without an activation barrier, with a reaction energy of 273.5~\kjmol, consistent with the gas-phase calculations of \citet{joshi2022chemical}. This difference likely reflects the difficulty of locating transition states for this type of reaction on the ice surface.

Finally, while HCN is less reactive than HNC against H-addition reactions in our network, one could ask whether reactions with other radicals might lead to a different conclusion. Indeed, although direct HCN + radical chemistry on interstellar ices remains largely unexplored, the study of \citet{Boland2024} still reinforces this selectivity: OH-addition pathways differ strongly for HCN and HNC and generally proceed more readily from HNC.

\subsection{Astrophysical implications}\label{sec:astro}

\begin{figure}[!htbp]
    \centering
    \includegraphics[width=0.5\textwidth]{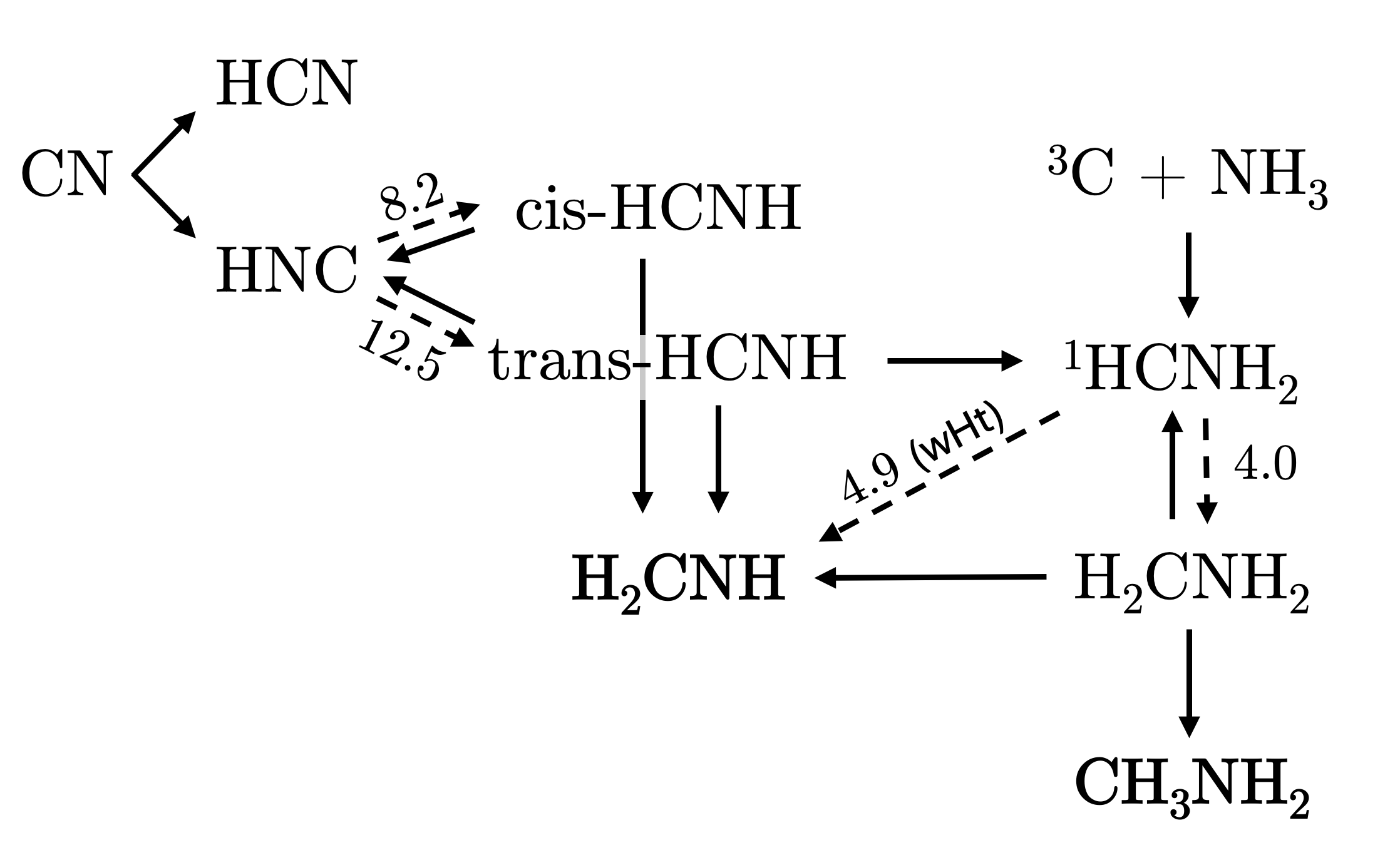}
    \caption{Simplified reaction network showing the main steps along the most efficient path in the formation of methanimine and methylamine. Unless specified, the arrows indicate hydrogenation or hydrogen abstraction steps. Solid lines indicate barrier-less channels, while dashed ones indicate the presence of an activation barrier. The entrance channel through \ce{^3C + NH3} involves a minimum of two steps; see the main network in Fig. \ref{fig:summary}.}
    \label{fig:concl_summary}
\end{figure}

Our results on the hydrogenation of HCN and HNC on interstellar ices reveal a network comprising both barrier-less and barrier-mediated steps, many of which are likely enhanced by quantum tunneling. A key outcome is that HCN is expected to survive longer than HNC under interstellar conditions, owing to the higher reactivity of HNC toward atomic hydrogen. For clarity, a simplified representation of the chemical network is provided in Fig.~\ref{fig:concl_summary}, which serves as a guide throughout the following discussion.

As summarized in Fig.~\ref{fig:concl_summary}, hydrogenation preferentially proceeds from \ce{HNC \to HC^.NH}. From there, the system can evolve toward \ce{^1[HC{:}NH2] \to H2C^.NH2 \to CH3NH2} or remain at the methanimine stage since the alternative route \ce{H2CNH \to H3CN^.H \to CH3NH2} is energetically less favorable. Starting from HNC, formation of \ce{HC^.NH} is associated with activation barriers of $\sim$8 and 13~\kjmol, depending on the conformation of the product. In addition, H-abstraction reactions can also occur, and in cases that involve a radical, they are barrier-less. This indicates that the relative orientation of the encounter between radicals and H atoms plays an important role in determining the reaction outcome, and suggests the presence of H-addition/H-abstraction loops within the network. Methanimine may act as a chemical sink, in the sense that one more step forward of backward in the network is mediated by a sizable activation energy barrier (of around 25~\kjmol), and therefore would require efficient quantum tunneling, while other intermediates such as \ce{^1[HC{:}NH2]} show a small barrier to the next step (4.0~\kjmol).
In addition, according to our calculations, wHt isomerizations between methanimine and \ce{^1[HC{:}NH2]} are only exothermic in the direction toward the formation of methanimine, with a low activation energy (4.9~\kjmol).

One may ask what the importance of reactions involving molecular hydrogen (\ce{H2}) is. Here we observe that not all \ce{\text{radical} + H2 \to \text{molecule} + H} are exoergic and therefore cannot take place under interstellar conditions. The only case where \ce{H2} participates as a reactant with a mild barrier is \ce{trans-HC^.NH + H2 \to H2CNH} (with a barrier of 36.6~\kjmol), still rather high. 
Therefore, the presence of free H atoms in the environment is rather important to fuel the network in Fig. \ref{fig:summary}. In any case, other channels exist that can lead to the formation of methanimine and methylamine, such as radical–radical chemistry \citep[e.g., \ce{CH3 + NH2}][]{Garrod2008,ER2022ApJS}, gas-phase processes \citep[e.g.,][]{Turner1999ApJ,Sil2018ApJ,Luthra2023}, or a mix of the two \citep[such as via H-abstraction from methylamine][]{joshi2022chemical}. Similarly, the further evolution of methanimine and methylamine is not limited to surface chemistry, but gas-phase chemistry can also play an important role \citep{Vazart2015ApJ,vazart2016state,lupi2020methanimine,deJesus2021}.

In addition to the final products of the hydrogenation network, there are intermediates that are worth exploring in more detail. \citet{SanAndres2023MNRAS} propose that \ce{H2CN^.}/\ce{C^.NH2} can be used as temperature tracers, much like HCN/HNC is routinely used \citep{Hacar2020}. Looking at our network and the one by \citet{molpeceres2024carbon}, \ce{H2CN^.} is much more likely to be produced from the hydrogenation of HCN, while \ce{C^.NH2} seems to be a product of an H-abstraction on \ce{^3[{:}CNH3]}, formed from \ce{^3C + NH3}. The isomerization between these two isomers seems rather ineffective, since wHt reactions favor the formation of trans-\ce{HC^.NH}, and in addition, they can easily be destroyed by H-additions or abstractions. In any case, we calculated their binding energy distributions following the same approach as in \citet[see our Fig.~\ref{fig:BE_H2CN_CNH2}]{enrique2024complex}, finding that, on average, \ce{C^.NH2} binds more strongly to the water ice matrix than \ce{H2CN^.}. The former shows a bimodal distribution with two peaks at around 42 and 58~\kjmol, while the latter sports a single peak at $\sim$22~\kjmol. In summary, these two species, if produced on interstellar ices, are quite likely to come from different direct parent species in our network.\\

Methanimine and methylamine are important intermediates in astrochemistry, as both are potential precursors to pre-biotically relevant molecules such as glycine (\ce{NH2CH2COOH}; \citealt{bernstein2002racemic, elsila2007mechanisms}). In any case, the \ce{H2C^.NH2} radical may play an important role in glycine formation through the radical–radical coupling with the \ce{HOC^.O} radical, as proposed in the reaction: \ce{HO^.CO + ^.CH2NH2 \to NH2CH2COOH} \citep{Woon2002}. Methanimine is both efficiently formed and relatively stable, and it is therefore expected to accumulate on interstellar ice surfaces. As a result, methanimine may serve as a key precursor to glycine formation. Although conversion to the \ce{H2C^.NH2} radical requires overcoming a moderately high activation barrier (22.9~\kjmol), the enhanced abundance of methanimine can still facilitate its involvement in downstream glycine-forming pathways.\\

Finally, we examined the effects of deuterium substitution along the main formation routes to methanimine and methylamine (see Sect.~\ref{sec:deuter}). Overall, D substitution has only a minor impact on activation energies but leads to lower imaginary frequencies, indicating slightly wider effective barriers and thus reduced tunneling efficiency for reactions involving direct D addition. As a result, H-addition pathways remain kinetically favored, although the overall structure of the network is not altered. From an astrochemical perspective, this implies that multiple deuteration pathways remain viable. In the case of methanimine, both DNC-derived routes and pathways involving \ce{NH2D} can contribute, while the weak sensitivity of wHt reactions to isotopic substitution suggests that exchange with the ice matrix can efficiently redistribute deuterium. Consequently, we do not expect that the presence of singly deuterated methanimine uniquely constrains its formation pathway. A similar situation applies to methylamine. Its formation through successive hydrogenation and abstraction steps allows multiple D-substitution patterns, with their relative abundances primarily governed by the local availability of deuterated precursor species rather than by a single dominant mechanism.
All of this leads to the conclusion that the most likely deuterated products are \ce{H2CND} (linked to DNC), \ce{HDCNH}, and \ce{CH3NHD} (linked with \ce{NH2D}) and \ce{CH2DNH2}.

\section{Conclusions}\label{sec:conclusions}

In this work we explored the hydrogenation network connecting HCN and HNC to methylamine on interstellar water ices. A schematic overview of the network is provided in Fig.~\ref{fig:concl_summary}. Our calculations show that HNC constitutes the most favorable entry point into the hydrogenation sequence, as H addition leading to cis-\ce{HC^.NH} proceeds over significantly lower barriers than the corresponding reactions involving HCN. Water-assisted HNC$\rightarrow$HCN isomerization is also unlikely to compete under cold pre-stellar conditions due to the high associated barriers. These results suggest that HCN can survive longer on interstellar ice surfaces than its isomer HNC.

Within the network, methanimine (\ce{H2CNH}) occupies a central position because several pathways converge toward this intermediate. However, its further hydrogenation requires overcoming moderate activation barriers, which may slow its conversion and effectively make it a chemical sink under interstellar conditions. The story is different for the two higher-energy isomers of methanimine. On the one hand, aminocarbene (\ce{^1[HC{:}NH2]}) provides an efficient route toward full hydrogenation. The conversion of this intermediate to \ce{H2C^.NH2} proceeds over a very small barrier (4.0~\kjmol), facilitating rapid progression toward methylamine, whereas pathways leading to triplet methyl nitrene (\ce{^3[H3CN{:}]}) are disfavored.

In any case, if the radical intermediates \ce{H3CN^.H} or \ce{H2C^.NH2} are formed, the final hydrogenation step leading to methylamine (\ce{CH3NH2}) is barrier-less, making the final saturation efficient under H-rich conditions. Conversely, H-abstraction from methylamine generally involves sizable barriers. Reactions involving molecular hydrogen are instead mostly inefficient, as they are associated with comparatively high activation barriers or are endoergic, highlighting the dominant role of atomic hydrogen in driving this chemistry during the early stages of star formation.

The local ice environment strongly influences reactivity. For instance, differences in binding energies can lead to less stable reactant configurations with lower relative activation barriers compared to cases where the adsorbate resides in a deeper binding site. Similarly, for wHt reactions, the structure of the hydrogen-bond network, such as the number of water molecules involved in the transference of the H atom, plays a critical role in determining the reaction energetics.
Finally, we have also investigated the effect of deuterium substitution, which mainly affects barrier widths associated with tunneling efficiencies, having only a limited impact on barrier heights.
Additional supporting information and optimized molecular geometries are available in the online repository \citep{zenodo_dataset}.

\begin{acknowledgements}
This project has received funding from the Horizon Europe Framework Programme (HORIZON) under the Marie Skłodowska-Curie grant agreement No 101149067, ``ICE-CN''. 
Finally, this work was granted access to the HPC resources of the high-performance computer SNELLIUS, part of the SURF cooperative of educational and research institutions in the Netherlands, under project No EINF-12426.
\end{acknowledgements}

\bibliographystyle{aa}
\bibliography{mybiblio}

\begin{appendix}

\onecolumn
\begingroup
\section{Reaction energetic descriptors}

\setlength{\LTleft}{0pt}
\setlength{\LTright}{0pt}
\begin{longtable}{@{}llcclc@{}}
\caption{Reaction energetic descriptors and comparison to the literature.}
\label{Table:summary_and_compare}\\
\toprule
Label & Reaction & $\Delta U_{rx}$ & $\Delta U^{\dagger}$ & $\Delta U^{\dagger}$ Other Works & T$_c$ (K) \\
\midrule
\endfirsthead
\caption{continued.}\\
\multicolumn{6}{c}{}\\
\toprule
Label & Reaction & $\Delta U_{rx}$ & $\Delta U^{\dagger}$ & $\Delta U^{\dagger}$ Other Works & T$_c$ (K) \\
\midrule
\endhead

\midrule
\multicolumn{6}{r}{Continued on next page}\\
\endfoot

\bottomrule
\multicolumn{6}{@{}p{\linewidth}@{}}{\footnotesize
$^{a}$ \citet{molpeceres2024carbon} (ice), $^{b}$ \citet{Woon2002} (gas), $^{c}$ \citet{Talbi1996} (gas), $^{d}$ \citet{Majumdar2018} (gas), $^{e}$ \citet{joshi2022chemical} (gas), $^{f}$ \citet{deJesus2021} (gas), $^{g}$ \citet{Ferrero2024} (ice), $^{h}$ \citet{baiano2022gliding}.}\\
\multicolumn{6}{@{}p{\linewidth}@{}}{\footnotesize
* Calculated with respect to the asymptote, omitting ZPE corrections.}\\
\multicolumn{6}{@{}p{\linewidth}@{}}{\footnotesize
** Calculated with $\omega$B97m-D3(BJ)/ma-def2-TZVP due to unrealistic ZPE correction at M062X-D3(0).}\\
\multicolumn{6}{@{}p{\linewidth}@{}}{\footnotesize
$^{\dagger}$ The reaction does not stop at \ce{H2CNH2 + H}, but it directly evolves to methylamine over a very high barrier.}\\
\multicolumn{6}{@{}p{\linewidth}@{}}{\footnotesize
Reaction~\ref{chem:CH3NH2+H__H2CNH2+H2} appears twice in the table because two different binding configurations of the reactants were identified, leading to distinct activation barriers (see Sect.~\ref{sec:discussion}).}\\
\multicolumn{6}{@{}p{\linewidth}@{}}{\footnotesize
Reaction~\ref{chem:HCN__HNC} is also listed twice, corresponding to water-assisted H-transfer (wHt) mechanisms involving three or two water molecules, respectively.}\\
\multicolumn{6}{@{}p{\linewidth}@{}}{\footnotesize
The reactions \ce{^3[{:}CNH3] + H^. \to HC^.NH3} and \ce{HC^.NH3 ->[wHt][] H2C^.NH2} (\rxnlabel{chem:CNH3__HCNH3}{R42} and \rxnlabel{chem:HCNH3__H2CNH2}{R43}) are not included in this table, as they ultimately lead to \ce{H2C^.NH2} after direct optimization of \ce{HC^.NH3} in our model.}
\endlastfoot

\multicolumn{6}{@{}l}{\textit{H-addition reactions}}\\
\midrule
\rxnlabel{chem:HCN+H__H2CN}{R1} & \ce{HCN + H^. \to H2CN^.} & -90.3 & 28.4 & 20.1$^a$, 30.5$^b$, 63.6$^c$, 28.3$^d$, 26.7$^f$ & 182 \\
\rxnlabel{chem:HCN+H__cis/trans-HCNH}{R2}-cis & \ce{HCN + H^. \to cis-HC^.NH} & -60.5 & 33.9 & 37.3$^a$, 53.5$^b$ & 261 \\
\rxnlabel{chem:HCN+H__cis/trans-HCNH}{R2}-trans & \ce{HCN + H^. \to trans-HC^.NH} & -78.5 & 41.7 & 74.4$^c$ & 280 \\
\rxnlabel{chem:HNC+H__CNH2}{R3} & \ce{HNC + H^. \to C^.NH2} & -43.7 & 55.9 & 55.6$^a$, 80.3$^c$ & 296 \\
\rxnlabel{chem:HNC+H__cis/trans-HCNH}{R4}-cis & \ce{HNC + H^. \to cis-HC^.NH} & -111.4 & 8.2 & 8.1$^a$ & 111 \\
\rxnlabel{chem:HNC+H__cis/trans-HCNH}{R4}-trans & \ce{HNC + H^. \to trans-HC^.NH} & -110.7 & 12.5 & 17.6$^c$ & 100 \\
\rxnlabel{chem:H2CN+H__H2CNH}{R5} & \ce{H2CN^. + H^. \to H2CNH} & -406.4* & BL & BL$^a$ & \\
\rxnlabel{chem:H2CN+H__tripH3CN}{R6} & \ce{H2CN^. + H^. \to ^3[H3CN{:}]} & -145.8 & 13.5 & 9.0$^a$ & 156 \\
\rxnlabel{chem:CNH2+H__singHCNH2}{R7} & \ce{C^.NH2 + H^. \to ^1[HC{:}NH2]} & -383.9* & BL & BL$^a$ & \\
\rxnlabel{chem:cis/trans-HNCH+H__H2CNH}{R8}-cis & \ce{cis-HC^.NH + H^. \to H2CNH} & -450.8* & BL & BL$^a$ & \\
\rxnlabel{chem:cis/trans-HNCH+H__H2CNH}{R8}-trans & \ce{trans-HC^.NH + H^. \to H2CNH} & -445.1* & BL &  & \\
\rxnlabel{chem:cis/trans-HNCH+H__singHCNH2}{R9}-cis & \ce{cis-HC^.NH + H^. \to ^1[HC{:}NH2]} & -328.7* & BL & BL$^a$ & \\
\rxnlabel{chem:cis/trans-HNCH+H__singHCNH2}{R9}-trans & \ce{trans-HC^.NH + H^. \to ^1[HC{:}NH2]} & -325.4* & BL &  & \\
\rxnlabel{chem:tripH3CN+H__H3CNH}{R10} & \ce{^3[H3CN{:}] + H^. \to H3CN^.H} & -385.1* & BL & BL$^a$ & \\
\rxnlabel{chem:H2CNH+H__H2CNH2}{R11} & \ce{H2CNH + H^. \to H2CN^.H2} & -145.9 & 22.9 & 19.4$^a$, 25.5$^b$, 12.7$^e$, 17.7$^f$ & 186 \\
\rxnlabel{chem:H2CNH+H__H3CNH}{R12} & \ce{H2CNH + H^. \to H3CN^.H} & -128.8 & 16.0 & 15.1$^a$, 19.2$^b$ & 171 \\
\rxnlabel{chem:singHCNH2+H__H2CNH2}{R13} & \ce{^1[HC{:}NH2] + H^. \to H2C^.NH2} & -270.1 & 4.0 & N/A$^a$ & 96 \\
\rxnlabel{chem:singHCNH2+H__HCNH3}{R14} & \ce{^1[HC{:}NH2] + H^. \to HC^.NH3} & -56.9 & 66.6 & 101.9 & 328 \\
\rxnlabel{chem:H3CNH+H__CH3NH2}{R15} & \ce{H3CN^.H + H^. \to CH3NH2} & -446.7* & BL & BL$^a$ & \\
\rxnlabel{chem:H2CNH2+H__CH3NH2}{R16} & \ce{H2C^.NH2 + H^. \to CH3NH2} & -426.4* & BL & BL$^a$ & \\

\midrule
\multicolumn{6}{@{}l}{\textit{H-abstraction and H$_2$-addition reactions}}\\
\midrule
\rxnlabel{chem:H2CN+H__HCN+H2}{R17} & \ce{H2CN^. + H^. \to HCN + H2} & -328.9 & BL & BL$^a$ & \\
\rxnlabel{chem:CNH2+H__HNC+H2}{R18} & \ce{C^.NH2 + H^. \to HNC + H2} & -383.8* & BL & BL$^a$ & \\
\rxnlabel{chem:cis/trans-HCNH+H__HCN+H2}{R19}-cis & \ce{cis-HC^.NH + H^. \to HCN + H2} & -367.7 & BL & BL$^a$ & \\
\rxnlabel{chem:cis/trans-HCNH+H__HCN+H2}{R19}-trans & \ce{trans-HC^.NH + H^. \to HCN + H2} & -351.1 & 2.2 &  & 162 \\
\rxnlabel{chem:cis/trans-HCNH+H__HNC+H2}{R20}-cis & \ce{cis-HC^.NH + H^. \to HNC + H2} & -335.8* & BL & BL$^a$ & \\
\rxnlabel{chem:cis/trans-HCNH+H__HNC+H2}{R20}-trans & \ce{trans-HC^.NH + H^. \to HNC + H2} & -320.7* & BL &  & \\
\rxnlabel{chem:tripH3CN+H__H2CN+H2}{R21} & \ce{^3[H3CN{:}] + H^. \to H2CN^. + H2} & -279.6 & 6.7 & BL$^a$ & 135 \\
\rxnlabel{chem:H2CNH+H__cis/trans-HCNH+H2}{R22}-cis & \ce{H2CNH + H^. \to cis-HC^.NH + H2} & -33.9 & 24.8 & 45.0$^a$ & 340 \\
\rxnlabel{chem:H2CNH+H__cis/trans-HCNH+H2}{R22}-trans & \ce{trans-HC^.NH + H2 \to H2CNH + H^.} & -24.7 & 36.6 &  & 371 \\
\rxnlabel{chem:H2CNH+H__H2CN+H2}{R23} & \ce{H2CNH + H^. \to H2CN^. + H2} & -54.0 & 29.4 &  & 406 \\
\rxnlabel{chem:singHCNH2+H__cis/trans-HCNH+H2}{R24}-cis & \ce{^1[HC{:}NH2] + H^. \to cis-HC^.NH + H2} & -138.5 & 28.3 & 22.7$^a$ & 349 \\
\rxnlabel{chem:singHCNH2+H__cis/trans-HCNH+H2}{R24}-trans & \ce{^1[HC{:}NH2] + H^. \to trans-HC^.NH + H2} & -140.7 & 14.7 & 22.7$^a$ & 349 \\
\rxnlabel{chem:singHCNH2+H__CNH2+H2}{R25} & \ce{^1[HC{:}NH2] + H^. \to C^.NH2 + H2} & -68.1** & 8.8** & 3.5$^a$ & 177 \\
\rxnlabel{chem:H3CNH+H__tripCH3N+H2}{R26} & \ce{H3CN^.H + H^. \to ^3[H3CN{:}] + H2} & -73.0 & 15.6 & 12.9$^a$ & 356 \\
\rxnlabel{chem:H3CNH+H__H2CNH+H2}{R27} & \ce{H3CN^.H + H^. \to H2CNH + H2} & -301.7 & 4.8 & 1.8$^a$ & 129 \\
\rxnlabel{chem:HCNH2+H__singHCNH2+H2}{R28} & \ce{H2C^.NH2 + H^. \to ^1[HC{:}NH2] + H2} & -154.4* & BL & N/A$^a$ & \\
\rxnlabel{chem:H2CNH2+H__H2CNH+H2}{R29} & \ce{H2C^.NH2 + H^. \to H2CNH + H2} & -273.5 & BL & N/A$^a$ & \\
\rxnlabel{chem:HCNH3+H__singHCNH2+H2}{R30} & \ce{HC^.NH3 + H^. \to ^1[HC{:}NH2] + H2} & -373.6 & BL &  & \\
\rxnlabel{chem:HCNH3+H__CNH3+H2}{R31} & \ce{HC^.NH3 + H^. \to ^3[{:}CNH3] + H2} & -52.7** & 6.6** &  & 57 \\
\rxnlabel{chem:CH3NH2+H__H2CNH2+H2}{R32} & \ce{CH3NH2 + H^. \to H2C^.NH2 + H2} & -67.3, -38.1 & 9.9, 30.2 & 37.3$^a$, 33.4$^e$ & 125, 337 \\
\rxnlabel{chem:CH3NH2+H__H3CNH+H2}{R33} & \ce{CH3NH2 + H^. \to H3CN^.H + H2} & -14.3 & 48.0 & 44.8$^a$, 39.7$^e$ & 414 \\
\rxnlabel{chem:HCNH2+H2__CH3NH2}{R34} & \ce{^1[HC{:}NH2] + H2 \to CH3NH2}$^{\dagger}$ & -231.7 & 102.6 &  & 270 \\

\midrule
\multicolumn{6}{@{}l}{\textit{Water-assisted H-transfer and isomerization reactions}}\\
\midrule
\rxnlabel{chem:HCN__HNC}{R35} & \ce{HNC ->[wHt][] HCN} & -32.4, -33.7  & 35.1, 73.6 & 49.6$^{h}$ & 221, 149 \\
\rxnlabel{chem:cis/trans-HCNH__H2CN}{R36} & \ce{trans-HC^.NH ->[wHt][] H2CN^.} & -32.9 & 83.7 &  & \\
\rxnlabel{chem:cis-HCNH__trans-HCNH}{R37}-isom & \ce{cis-HC^.NH ->[isom][] trans-HC^.NH} & -9.6 & 23.9 &  & 136 \\
\rxnlabel{chem:cis-HCNH__trans-HCNH}{R37}-wHt & \ce{cis-HC^.NH ->[wHt][] trans-HC^.NH} & -22.6 & 65.5 &  & 196 \\
\rxnlabel{chem:CNH2__cis/trans-HCNH}{R38} & \ce{C^.NH2 ->[wHt][] trans-HC^.NH} & -47.1 & 30.8 & 88.8$^a$ & 52 \\
\rxnlabel{chem:HCNH2__H2CNH}{R39} & \ce{^1[HC{:}NH2] ->[wHt][] H2CNH} & -106.1 & 4.9 & 22.7$^a$, 14.0$^g$ & 187 \\
\rxnlabel{chem:H3CNH__H2CNH2}{R40} & \ce{H3CN^.H ->[wHt][] H2C^.NH2} & -21.5 & 69.3 &  & 100 \\
\rxnlabel{chem:HCNH3__H2CNH2}{R41} & \ce{HC^.NH3 ->[wHt][] H2C^.NH2} & -209.3 & BL & BL$^a$ & \\
\end{longtable}
\tablefoot{
Summary of the reactions considered in this work and their energetics.
Each reaction is assigned a label (R1, R2, ...) that is used throughout the text.
Energy units are in \kjmol. $\Delta U_{rx}$ and $\Delta U^{\ddagger}$ are the reaction and activation energies of each reaction, $T_c$ is the tunneling crossover temperature. BL stands for barrier-less and N/A for ``not an answer.''
}
\endgroup


\section{BE distributions of CNH$_2$ and CH$_2$N}\label{apx_sec:BE_CNH2_CH2N}

\begin{figure}[!htbp]
    \centering
    \includegraphics[width=0.8\linewidth]{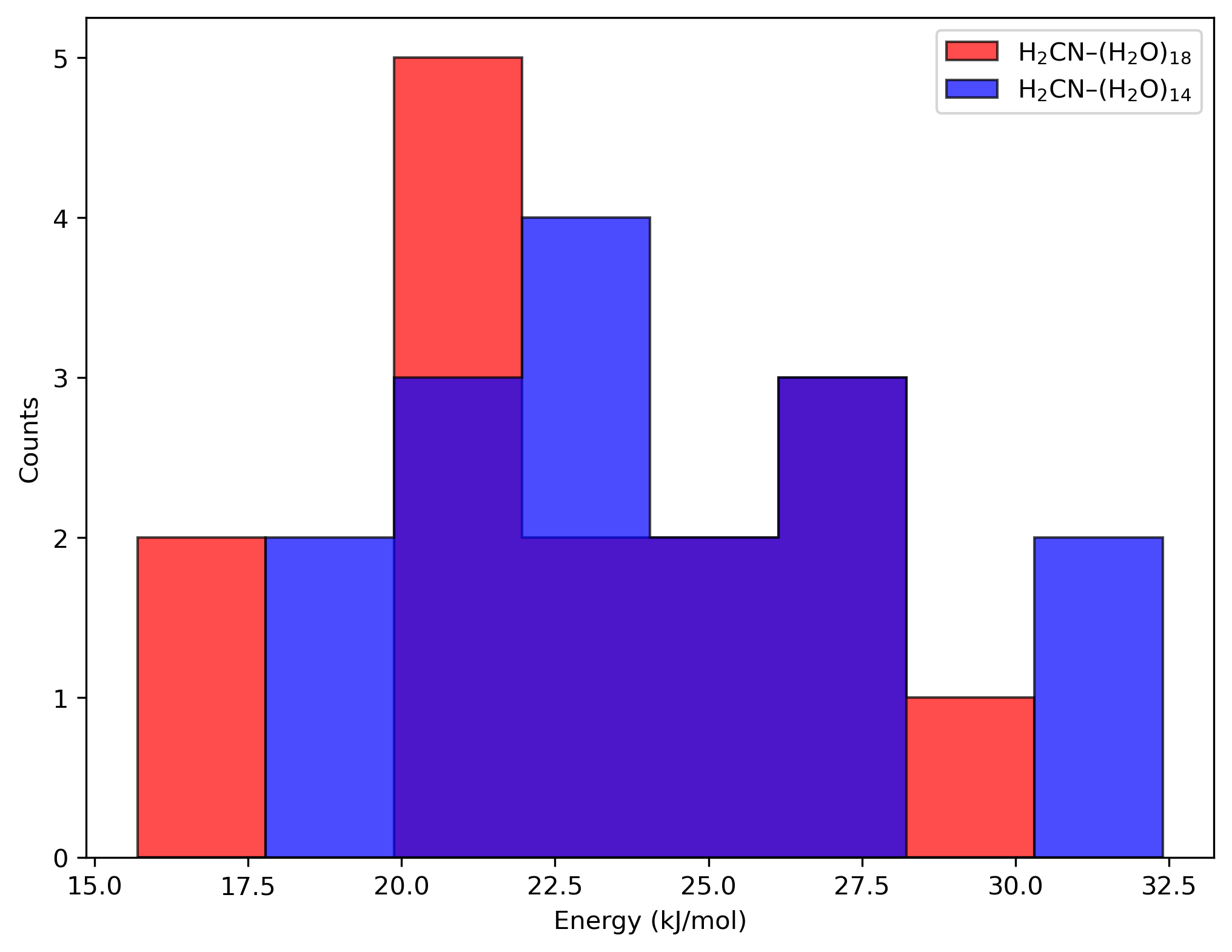}
    \includegraphics[width=0.8\linewidth]{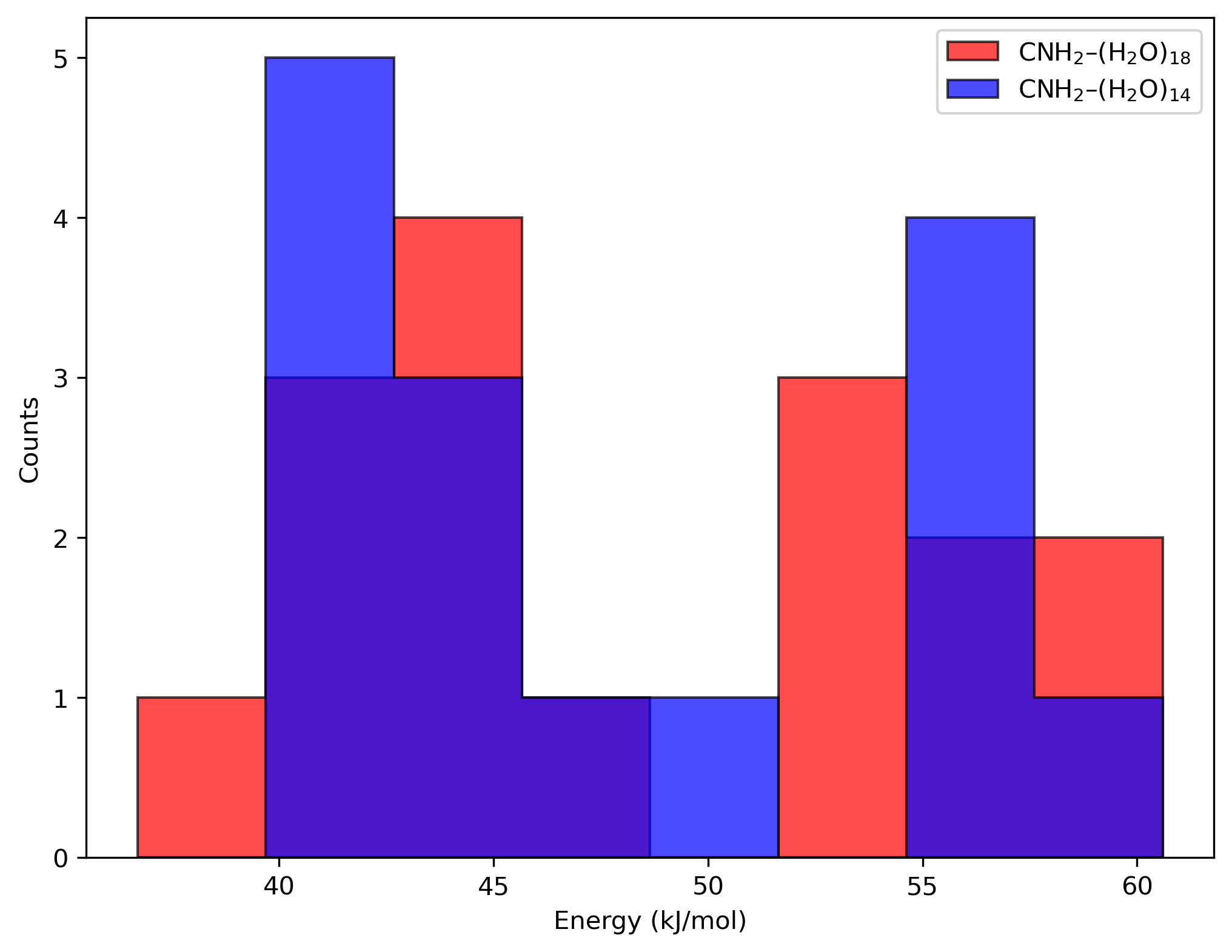}
    \caption{Binding energy distributions for \ce{H2CN} (left) and \ce{CNH2} (right) on 14 and 18 water ice molecule clusters.}
    \label{fig:BE_H2CN_CNH2}
\end{figure}

\begin{figure*}[!htbp]
    \centering
    \includegraphics[width=0.25\textwidth]{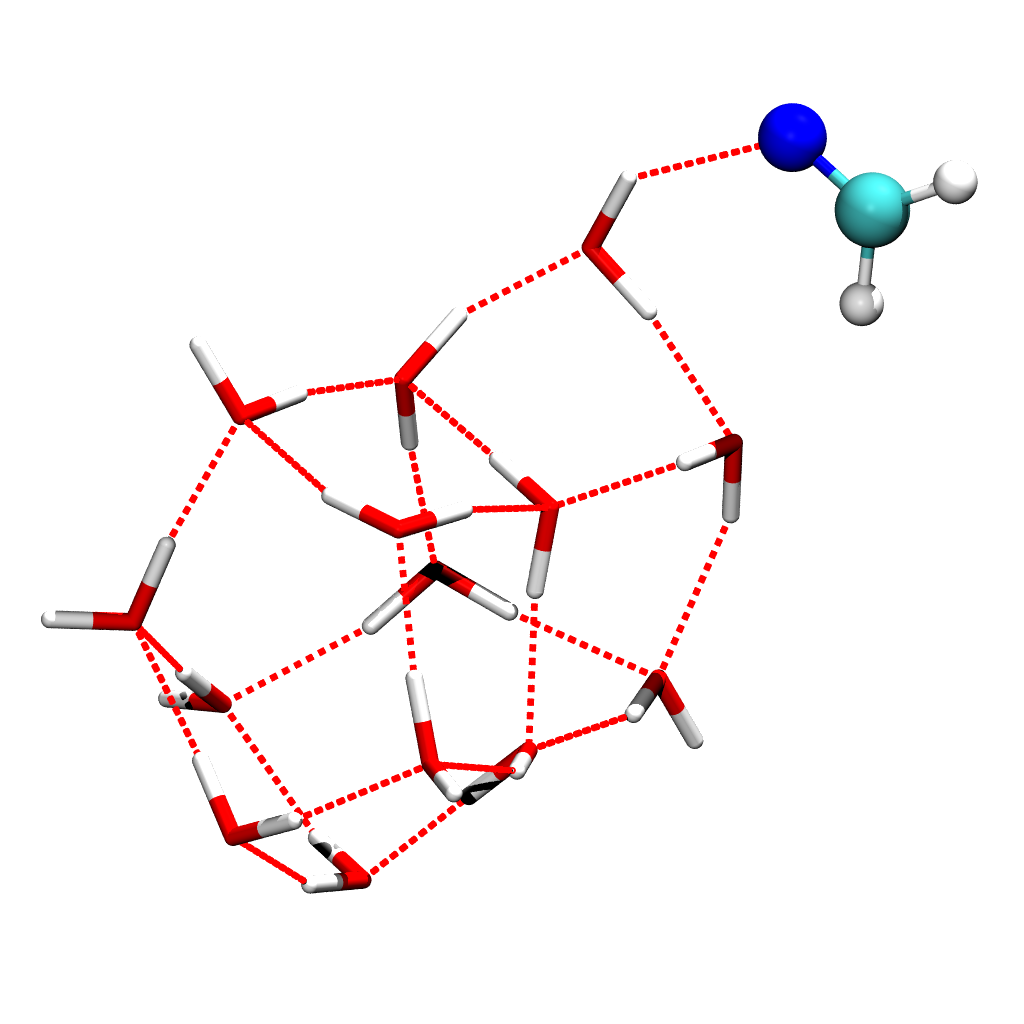}
    \includegraphics[width=0.25\textwidth]{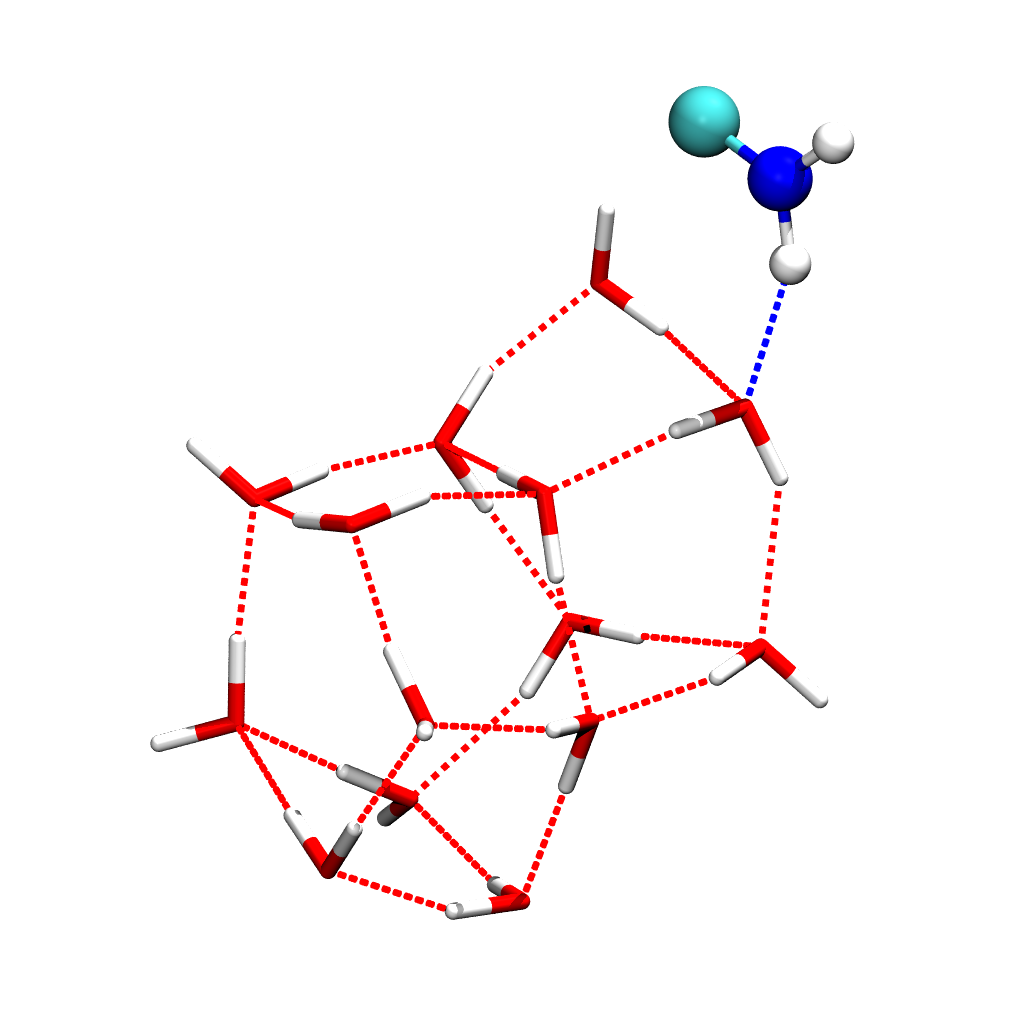}
    \includegraphics[width=0.25\textwidth]{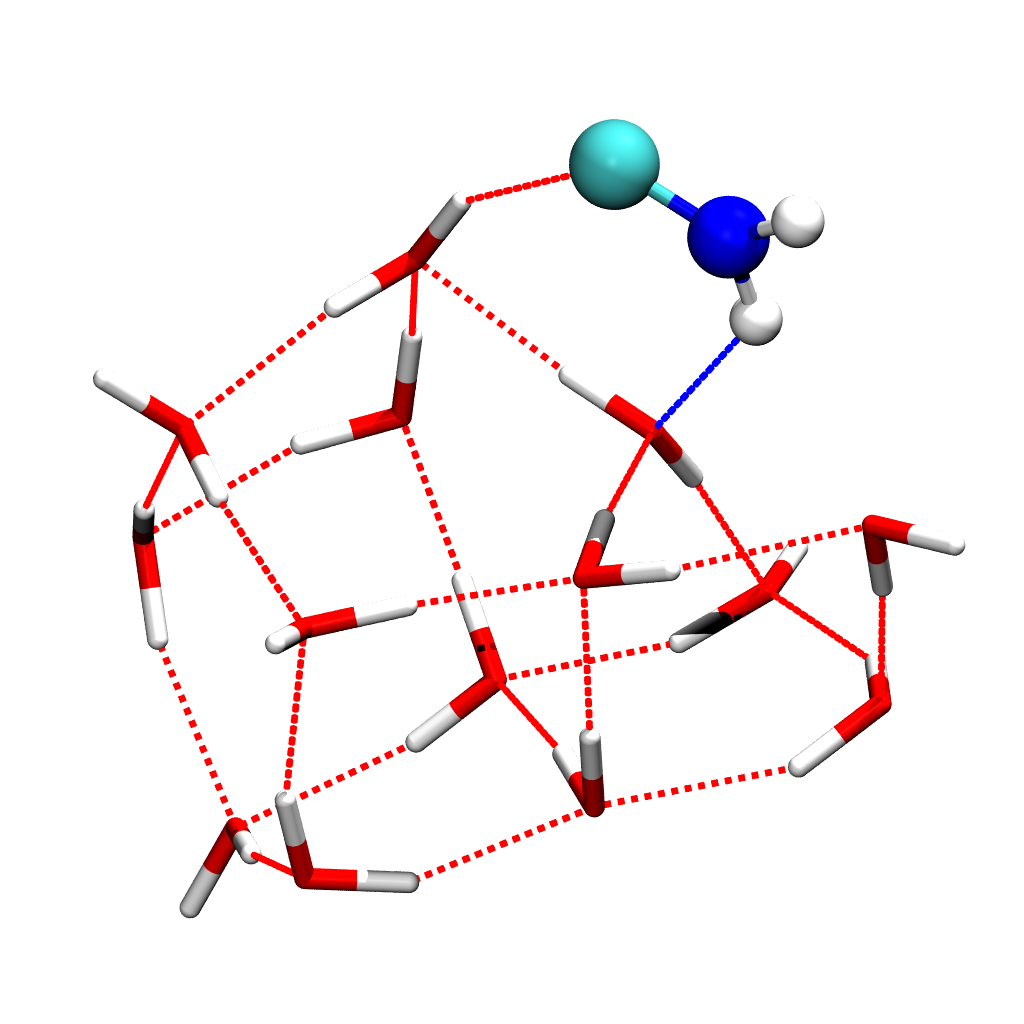}
    \caption{Geometries of the most frequent binding energies in Fig. \ref{fig:BE_H2CN_CNH2}. From left to right: \ce{H2CN} (22.1~\kjmol), \ce{CNH2} (42.5 and 55.9~\kjmol), and~\ce{(H2O)_{14}}.}
    \label{fig:BE_cases_H2CN_CNH2}
\end{figure*}

The binding energy distributions for \ce{H2CN^.} and \ce{C^.NH2} are shown in Fig.~\ref{fig:BE_H2CN_CNH2}. Representative examples of the most common interaction modes are presented in Fig.~\ref{fig:BE_cases_H2CN_CNH2}, including one case for \ce{H2CN} and two for \ce{CNH2}. The first two structures correspond to adsorption on the same binding site, illustrating how small structural differences, specifically the position of the two H atoms in the admolecule, can lead to significant differences in binding energy. The third case corresponds to a deep binding site for \ce{CNH2}, where the \ce{NH2} group is embedded within a small pocket-like structure in the ice, resulting in a substantially stronger interaction.

\section{Effect of the binding mode on activation energy barriers}

\begin{figure*}[!htbp]
    \centering
    \includegraphics[width=0.7\textwidth]{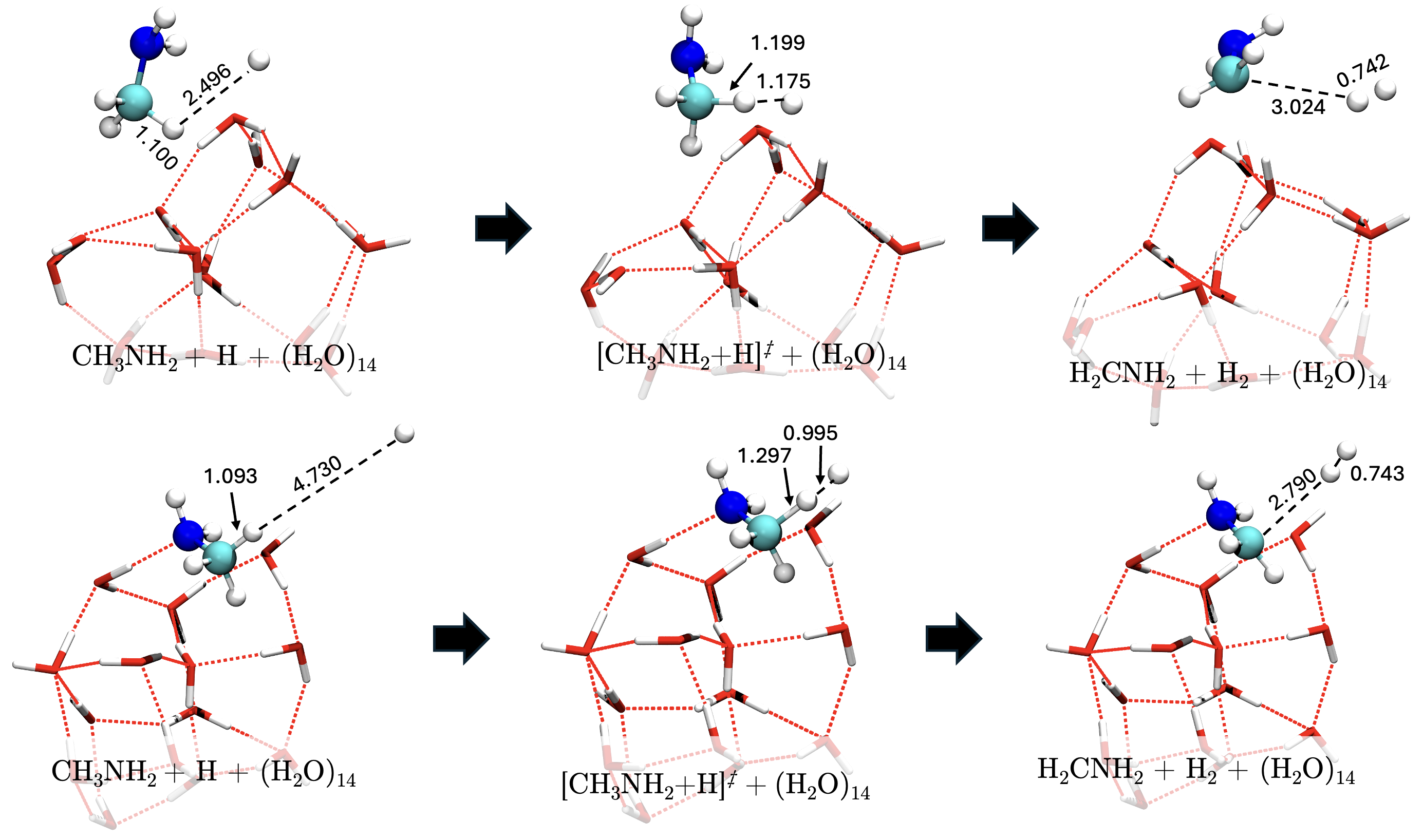}
    \caption{Stationary points along the potential energy surface for the H-abstraction reaction \ce{CH3NH2 + H^. -> H2C^.NH2 + H2}. The upper panel shows the configuration where the incoming H atom is located beneath the \ce{-NH2} group, weakening its interaction with the ice surface and leading to a lower activation barrier (9.9~\kjmol). The lower panel shows a configuration in which the H atom originates from a different binding site and does not perturb the interaction of \ce{CH3NH2} with the surface, resulting in a higher barrier (30.2~\kjmol). Bond distances are in \AA.}
    \label{fig:CH3NH2+H}
\end{figure*}

To illustrate the influence of the local binding configuration on the activation barrier, Fig.~\ref{fig:CH3NH2+H} shows the stationary points corresponding to reaction~\ref{chem:CH3NH2+H__H2CNH2+H2} discussed in Sect.~\ref{sec:discussion}. Depending on the relative position of the incoming H atom with respect to the \ce{-NH2} group of methylamine, two different activation barriers are obtained. When the H atom resides directly beneath the \ce{-NH2} group, the interaction of the amine moiety with the ice surface is weakened, destabilizing the reactant complex and resulting in a lower barrier (9.9~\kjmol). In contrast, when the H atom originates from a different binding site on the surface and does not perturb the interaction of \ce{CH3NH2} with the ice, the reactant configuration remains more stable, and the activation barrier increases to 30.2~\kjmol.

\section{Benchmark data}

A selection of 13 reactions was chosen to run a benchmark study. Table  \ref{tab:ref_vals_benchmark} contains their relative energetics (barriers and reaction energies) at the reference level. The relative differences of their activation energy with different functionals (in absolute value) can be found in \ref{tab:benchmark}. Raw energies can be obtained in the Zenodo repository of this project, \citet{zenodo_dataset}.
 
\begin{table}[!htbp]
\centering
\caption{Benchmark reference values.}
\label{tab:ref_vals_benchmark}
\begin{tabular}{lcc}
\hline
                                       & $\Delta E^{^\ddagger}$ & $\Delta E^{r}$ \\ \hline
A & 24.6  & -153.0         \\
B & 40.2  & -114.6         \\
C & 9.6   & -145.4         \\
D & 54.2  & -121.4         \\
E & 43.8  & -41.0          \\
F & 46.5  & -13.0          \\
G & 56.4  & -5.0           \\
H & 14.9  & -177.6         \\
I & 14.8  & -82.3          \\
J & 42.1  & -25.7          \\
K & 84.8  & -65.1          \\
L & 128.4 & -25.9          \\
M & 37.2  & -116.4         \\ \hline
\end{tabular}%
\tablefoot{
Energetics obtained with the reference method CCSD(T)/aug-cc-pVTZ//M062X-D3(0)/ma-def2-TZVP. Values in \kjmol. The considered reactions are A: \ce{HCN + H^. \to H2CN^.}, B: \ce{HCN + H^. \to HC^.NH}, C: \ce{HNC + H^. \to HC^.NH}, D: \ce{HNC + H^. \to C^.NH2}, E: \ce{CH3NH2 + H^. \to H2C^.NH2 + H2}, F: \ce{CH3NH2 + H \to CH3NH^. + H2}, G: \ce{H2CNH + H^. \to cis-HC^.NH + H2}, H: \ce{H2CNH + H^. \to H2C^.NH2}, I: \ce{H3CN^.H + H^. \to ^3[H3CN{:}] + H2}, J: \ce{trans-HC^.NH + H2 \to H2CNH + H^.}, K: wHt \ce{C^.NH2 \to trans-HC^.NH}, L: wHt \ce{H2C^.NH2 \to H3CN^.H} and M: wHt \ce{^1[HC{:}NH2] \to CH2NH}.}
\end{table}

\begin{table}[!htbp]
\centering
\caption{Benchmark energetic descriptors.}
\label{tab:benchmark}
\begin{tabular}{lccccccc}
\hline
                                        & 1 & 2 & 3 & 4 & 5 & 6 & 7 \\ \hline
A   & 8.4        & 0.7         & 3.3           & 1.4         & 6.4       & 0.3          & 0.4     \\
B   & 9.7        & 0.3         & 5.1           & 0.1         & 11.3      & 2.5          & 2.1     \\
C   & 5.3        & 3.3         & 1.4           & 1.4         & 4.5       & 1.9          & 2.6     \\
D   & 8.0        & 3.4         & 1.9           & 4.3         & 8.8       & 0.8          & 0.2     \\
E   & 12.6       & 0.2         & 6.4           & 4.8         & 16.1      & 7.2          & 6.1     \\
F   & 10.7       & 4.3         & 0.8           & 4.7         & 13.1      & 3.7          & 2.4     \\
G   & 12.4       & 2.2         & 7.2           & 8.3         & 17.6      & 8.3          & 7.3     \\
H   & 7.2        & 0.9         & 2.1           & 0.5         & 7.3       & 1.2          & 0.3     \\
I   & 5.8        & 2.1         & 1.3           & 4.1         & 8.9       & 0.4          & 0.6     \\
J   & 10.2       & 0.3         & 5.7           & 5.2         & 16.6      & 6.5          & 5.3     \\
K   & 13.4       & 16.7        & 0.3           & 6.7         & 9.7       & 4.5          & 6.8     \\
L   & 15.7       & 27.5        & 24.8          & 8.9         & 22.5      & 12.6         & 15.6    \\
M   & 4.4        & 11.5        & 5.1           & 4.3         & 3.9       & 0.5          & 1.6     \\ \hline
Avg (all)                           & 9.5        & 5.6         & 5.0           & 4.2         & 11.3      & 3.9          & 3.9     \\
Avg (wHt)                           & 11.2       & 18.6        & 10.1          & 6.7         & 12.0      & 5.9          & 8.0     \\
Avg (no wHt)                        & 9.0        & 1.8         & 3.5           & 3.5         & 11.0      & 3.3          & 2.7     \\ \hline
\end{tabular}%
\tablefoot{
Energy differences in activation barriers ($\Delta E^{\ddagger}$) between the tested density functional approximations (DFAs) and the reference values for the reactions reported in Table~\ref{tab:ref_vals_benchmark} (in \kjmol). Considered DFAs 1: BH\&HLYP-D4, 2: M06-2X-D3(0), 3: MPWB1K-D3(BJ), 4: M08HX-D3(0), 5: PW6B95-D4, 6: $\omega$B97m-D3(BJ), and 7: $\omega$B97m-V.}
\end{table}

\end{appendix}

\end{document}